\documentclass[prd,aps,showpacs,
11pt,
nofootinbib,%
tightenlines,%
eqsecnum%
]{revtex4-1}

\usepackage{amsmath,amssymb}
\usepackage{hyperref}
\usepackage{upgreek}

\usepackage{xcolor}

\newcommand{\be}{\begin{equation}}
\newcommand{\ee}{\end{equation}}
\newcommand{\bea}{\begin{eqnarray}}
\newcommand{\eea}{\end{eqnarray}}
\newcommand{\ba}{\begin{align}}
\newcommand{\ea}{\end{align}}
\newcommand{\nn}{\nonumber}

\newcommand{\RSS}{R_{\mathrm{D4}}}
\newcommand{\MKK}{M_{\mathrm{KK}}}

\newcommand{\gYM}{g_{\mathrm{YM}}}

\def\d{\text{d}}
\def\kk{\text{KK}}
\def\tr{\,\text{tr}\,}
\def\Tr{\,\text{Tr}\,}

\def\V{\mathcal{V}}

\def\xtau{\tau}

\def\pion{\Pi}
\def\Nc{N_c}

\def\q#1{q_{#1}}

\def\BPR{\cite{Brunner:2015oqa}}

\newif\ifmynotes
\mynotestrue

\definecolor{notetext}{rgb}{1,0,0}

\begin{document}

\title{Radiative Meson and Glueball Decays in the Witten-Sakai-Sugimoto Model}
\author{Florian Hechenberger}
\author{Josef Leutgeb}
\author{Anton Rebhan}
\affiliation{Institut f\"ur Theoretische Physik, Technische Universit\"at Wien,
        Wiedner Hauptstrasse 8-10, A-1040 Vienna, Austria}

\date{\today}
\begin{abstract}
We calculate radiative decay rates of mesons and glueballs in the top-down holographic Witten-Sakai-Sugimoto model with finite quark masses. After assessing to what extent this model agrees or disagrees with experimental data, we present its predictions for so far undetermined decay channels. Contrary to widespread expectations, we obtain sizeable two-photon widths of scalar, tensor, and pseudoscalar glueballs, suggesting in particular that the observed two-photon rate of the glueball candidate $f_0(1710)$ is not too large to permit a glueball interpretation, but could be even much higher. We also discuss the so-called exotic scalar glueball, which in the Witten-Sakai-Sugimoto model is too broad to match either of the main glueball candidates $f_0(1500)$ and $f_0(1710)$, but might be of interest with regard to the alternative scenario of the so-called fragmented scalar glueball. Employing the exotic scalar glueball for the latter, much smaller two-photon rates are predicted for the ground-state glueball despite a larger total width; relatively large two-photon rates would then apply to the excited scalar glueball described by the predominantly dilatonic scalar glueball. In either case, the resulting contributions to the muon $g-2$ from hadronic light-by-light scattering involving glueball exchanges are small compared to other single meson exchanges, of the order of $\lesssim 10^{-12}$.
\end{abstract}
\maketitle  

\tableofcontents

\section{Introduction and Summary}

Glueballs, bound states of gluons without valence quarks, have been proposed as a consequence of QCD  from the start \cite{Fritzsch:1972jv,Fritzsch:1973pi,Fritzsch:1975tx,Jaffe:1975fd}, but it is still a widely open question how they manifest themselves in the hadron spectrum \cite{Klempt:2007cp,Crede:2008vw,Ochs:2013gi,Klempt:2021nuf,Chen:2022asf}.
Lattice QCD \cite{Bali:1993fb,Morningstar:1999rf,Chen:2005mg,Gregory:2012hu,Chen:2021dvn}, mostly in the quenched approximation, provides more or less clear predictions for the spectrum, with a lightest glueball being a scalar, followed by a tensor glueball with an important role as the lightest state associated with the pomeron \cite{Donnachie:2002en}, a pseudoscalar glueball participating in the manifestation of the U(1)$_A$ anomaly responsible for the large mass of the $\eta'$ meson \cite{Rosenzweig:1981cu}, and towers of states with arbitrary integer spin as well as parity. However, it has turned out to be difficult to discriminate glueball states from bound states of quarks with the same quantum numbers with which they can mix, since the various available phenomenological models give strongly divergent pictures, in particular for the lightest glueballs. For the ground-state scalar glueball, the initially favored scenario that the isoscalar meson $f_0(1500)$ contains the most glue content while being strongly mixed with quarkonia \cite{Amsler:1995td,Close:2001ga,Close:2005vf}
is contested by models which identify the $f_0(1710)$
as a glueball candidate \cite{Lee:1999kv,Janowski:2014ppa,Cheng:2015iaa}
with more dominant glue content. The latter also appears favored by its larger production rate in supposedly gluon-rich radiative $J/\psi$ decays
\cite{Gui:2012gx}, but there it was proposed that the glue content might rather be distributed over several scalars involving a new meson $f_0(1770)$ previously lumped together with the established $f_0(1710)$ \cite{Sarantsev:2021ein,Klempt:2021nuf,Klempt:2021wpg}.

In order to clarify the situation, dynamical information on decay patterns is required from first principles, which is difficult to extract from Euclidean lattice QCD. Analytical approaches always involve uncontrollable approximations, albeit recently interesting progress has been made using Schwinger-Dyson equations \cite{Huber:2020ngt}. 

In this work we continue the analytical explorations made using gauge/gravity duality, which has been employed for studying glueball spectra in strongly coupled nonabelian theories shortly after the discovery of the AdS/CFT correspondence
\cite{Gross:1998gk,Csaki:1998qr,deMelloKoch:1998vqw,Hashimoto:1998if,Csaki:1999vb}, inspiring phenomenological ``bottom-up'' model building for glueball physics \cite{Boschi-Filho:2002wdj,Colangelo:2007pt,Forkel:2007ru,Li:2013oda,FolcoCapossoli:2015jnm,Ballon-Bayona:2017sxa,Rinaldi:2021dxh}.
Of particular interest here is the top-down construction of a dual to low-energy QCD in the large-$N_c$ limit from type-IIA string theory by Witten \cite{Witten:1998zw}, where the glueball spectrum has been obtained in \cite{Constable:1999gb,Brower:2000rp}. Sakai and Sugimoto \cite{Sakai:2004cn,Sakai:2005yt} have extended this model by a D-brane construction introducing $N_f$ chiral quarks in the 't Hooft limit $N_c\gg N_f$, which turns out to reproduce many features of low-energy QCD and chiral effective theory, not only qualitatively, but often semi-quantitatively, while having a minimal set of free parameters.

Glueball decay patterns have been first studied in the Witten-Sakai-Sugimoto (WSS) model for the scalar glueball in \cite{Hashimoto:2007ze} and revisited and extended in \cite{Brunner:2015oqa}. This involves a so-called exotic scalar glueball \cite{Constable:1999gb} for which it is unclear whether it should be identified with the ground-state glueball in QCD or instead be discarded together with the other states that more evidently do not relate to states in QCD. 

Assuming that the ground-state scalar glueball corresponds to the predominantly dilatonic bulk metric fluctuations which do not involve polarizations in the extra Kaluza-Klein dimension employed for supersymmetry breaking,
\cite{Brunner:2015yha,Brunner:2015oga} found that the resulting decay pattern could match remarkably well the one of the $f_0(1710)$ meson  when effects of finite quark masses are included (or $f_0(1770)$ when this is split off from a tetraquark $f_0(1710)$ \cite{Sarantsev:2021ein}). Instead of the chiral suppression postulated for flavor asymmetries of scalar glueball decay \cite{Chanowitz:2005du}, a nonchiral enhancement of decays into heavier pseudoscalars was obtained, which is correlated with a reduction of the $\eta\eta'$ decay mode \cite{Brunner:2015oga}.
This mechanism of flavor symmetry violation is absent for the tensor glueball, whose hadronic decays have been worked out also in 
\cite{Brunner:2015oqa}; hadronic decays of pseudoscalar and pseudovector glueballs have been studied in \cite{Brunner:2016ygk,Leutgeb:2019lqu,Brunner:2018wbv}.

In the present paper, we revisit and extend the study of glueball decay patterns
of \cite{Brunner:2015oqa,Brunner:2015yha,Brunner:2015oga} to also include radiative decays. As discussed already in \cite{Sakai:2005yt}, the WSS model
naturally incorporates vector meson dominance (VMD), crucially involving an infinite tower of vector mesons.
After assessing the predictions of the WSS model with regard to radiative decays of ordinary pseudoscalar and (axial) vector mesons, we analyze its corresponding results for glueballs.

Contrary to widespread expectations, the WSS model predicts that glueballs can have sizeable radiative decay widths in the keV range, exceeding even the claimed observation of two-photon rates for $f_0(1710)$ by the BESIII collaboration \cite{Belle:2013eck}, which was taken as evidence against its glueball nature.

In this context we also reconsider the exotic scalar glueball, which differs from the dilatonic one in that it has smaller couplings to vector mesons as well as photons, while having a total width in excess of the one of either $f_0(1500)$ or $f_0(1710)$, when its mass is suitably adjusted. As such it may instead be a candidate for the so-called fragmented scalar glueball proposed in \cite{Sarantsev:2021ein,Klempt:2021nuf,Klempt:2021wpg}, which is a wider resonance distributed over $f_0(1710)$, a novel $f_0(1770)$, $f_0(2020)$, and $f_0(2100)$, without showing up as an identifiable meson on its own.

In the case of the tensor glueball, where the WSS model is unequivocal in identifying the ground state, even though its mass also needs correction,
we find again two-photon rates in the keV region, larger than the old predictions of Kada et al.\ \cite{Kada:1988rs}, but comparable to those obtained by Cotanch and Williams \cite{Cotanch:2005ja} using VMD. (The latter have obtained even larger two-photon rates for the scalar glueball, which are an order of magnitude above the WSS results.)

The next heavier glueball, the pseudoscalar glueball, which plays an important rule in the realization of the U(1)$_A$ anomaly \cite{Leutgeb:2019lqu}, is also found to have two-photon rates in the keV region. 

Because of their sizeable two-photon coupling in the WSS model, we consider also the effect the lightest three glueballs may have as single-meson contributions to hadronic light-by-light scattering, which is an important ingredient of the Standard Model prediction of the anomalous magnetic moment of the muon \cite{Aoyama:2020ynm} $a_\upmu=(g-2)_\upmu/2$.
With the dilatonic scalar glueball as ground state, we find results
of $a_\upmu^{G}=-(1\dots1.6)\times 10^{-12}$, and one order of magnitude smaller when the exotic scalar glueball is used instead with mass raised to the value of the fragmented glueball of \cite{Sarantsev:2021ein}.
With its larger mass and comparable two-photon rate, the tensor glueball
is bound to contribute less than the dilatonic scalar glueball.
The pseudoscalar glueball, which contributes with a different sign, yields
$a_\upmu^{G_{PS}}=+(0.2\dots0.4)\times 10^{-12}$ depending on its actual mass.
All these results are thus safely smaller than the current uncertainties
in the hadronic light-by-light scattering contributions to $a_\upmu$.

\section{The Witten-Sakai-Sugimoto model augmented by quark masses}

The Witten-Sakai-Sugimoto (WSS) model \cite{Sakai:2004cn,Sakai:2005yt} is constructed
by placing a stack of $N_f$ flavor probe 
D8 and $\overline{\text{D8}}$-branes into the
near-horizon double Wick rotated black D4-brane background proposed
in \cite{Witten:1998zw} as a supergravity dual of four-dimensional
$\mathrm{U}(\Nc\to\infty)$ Yang-Mills (YM) theory at low energies, where supersymmetry and conformal symmetry are broken by compactifications.
It thus serves as a model for the low-energy limit of large $N_c$ QCD
with $N_f\ll N_c$, corresponding to a quenched approximation when extrapolated to $N_f=N_c=3$.
The background geometry is given by the metric
\begin{align}
  & \d s^2 = \left( \frac{U}{\RSS} \right)^{3/2} \left[\eta_{\mu\nu} \d x^\mu \d x^\nu + f(U) \d \xtau^2\right]+\left(\frac{\RSS}{U}\right)^{3/2} \left[\frac{\d U^{2}}{f(U)} + U^2 \d\Omega_4^2\right],\nonumber \\
  & e^{\phi} = g_{s}\left(\frac{U}{\RSS}\right)^{3/4},\qquad F_{4}=\d C_{3}=\frac{(2\pi l_s)^3\Nc}{V_{4}}\epsilon_{4},\qquad f(U)=1-\frac{U_\kk^3}{U^{3}},\label{eq:background}
\end{align}    
with dilaton $\phi$ and Ramond-Ramond three-form field $C_{3}$, a solution
of type IIA supergravity, whose bosonic part of the action reads 
\begin{widetext}
\begin{equation}
S_{\text{grav}}= \frac{1}{2\kappa_{10}^2}\int\d^{10}x\sqrt{-g}\left[e^{-2\phi}\left(R+4\left(\nabla\phi\right)^{2}\right)-\frac12 |F_{4}|^{2}\right].\label{eq:10dAction}
\end{equation}    
\end{widetext}

The $N_c$ D4-branes extend along the directions parametrized by the coordinates
$x^{\mu}$, $\mu=0,1,2,3$ and another spatial dimension with coordinate $\xtau$, while $U$ corresponds to the
radial (holographic) direction transverse to the D4-brane. The remaining four transverse
coordinates span a unit $S^{4}$ with line element $\d\Omega_{4}^{2}$,
volume form $\epsilon_{4}$ and volume $V_{4}=8\pi^{2}/3$. The $\xtau$-direction
is compactified to a supersymmetry breaking $S^{1}$, whose period
is chosen as 
\begin{align}
\xtau\simeq \xtau+\delta\xtau=\xtau+2\pi M_{\kk}^{-1},\quad & M_{\kk}=\frac{3}{2}\frac{U_\kk^{1/2}}{\RSS^{3/2}},
\end{align}
to avoid a conical singularity at $U=U_{\kk}$. The radius
$\RSS$ is related to the string coupling $g_{s}$ and the string length
$l_{s}$ through $\RSS^{3}=\pi g_{s}N_{c}l_{s}^{3},$ and the 't Hooft coupling of the dual four-dimensional Yang-Mills theory is given by
\be\label{gYMNc}
\lambda=\gYM^2 N_c=\frac{g_5^2}{\delta\xtau}N_c=2\pi g_s l_s\MKK N_c.
\ee

The flavor D8 and $\overline{\text{D8}}$-branes extend along $x^{\mu}$,
$U$, and the $S^{4}$. They are placed antipodally on the $\xtau$-circle
to join at $U_\kk$. In adopting the probe approximation, i.e. $\Nc\gg N_{f}$
for the $N_{f}$ D8 branes, one can ignore backreactions from the D8-branes
to the D4-brane background. The gauge fields on the D8-branes, which are dual to left and right chiral quark currents separated in the Kaluza-Klein ($\tau$) direction, are
governed at leading order by a Dirac-Born-Infeld (DBI) plus Chern-Simons
(CS)  action
\begin{align}
S_{\text{DBI}}= & -T_{8}\int\d^{9}x e^{-\phi}\Tr \sqrt{-\det\left(g_{MN}+2\pi\alpha^{\prime}F_{MN}\right)},\nonumber \\
S_{\text{CS}}= & T_{8}\int_{D8}C\wedge\Tr\left[\exp\left\{ \frac{F}{2\pi}\right\} \right]\sqrt{\hat{A}(\mathcal{R})},\label{eq:9dAction}
\end{align}
where $\hat{A}(\mathcal{R})$ is the so-called A-roof genus \cite{Green:1996dd,Polchinski1998}.

Considering only SO(5)-invariant excitations and restricting to terms quadratic in the field strength, the nine-dimensional DBI action can be reduced to a five-dimensional Yang-Mills theory with action \cite{Sakai:2004cn,Sakai:2005yt}\footnote{Note that in \eqref{eq:5dDBI} one uses the Minkowski metric $\eta_{\mu\nu}$, in the mostly plus
convention, to contract the four-dimensional spacetime indices.}
\begin{equation}
S_\text{D8}^\text{DBI} = -\kappa \int  \d^4x\, \d z\, {\Tr}\left[ \frac{1}{2} K^{-1/3} F_{\mu\nu}^2 + M_\kk ^2 K F_{\mu z}^2 \right],\label{eq:5dDBI}
\end{equation}
with
\be
\kappa\equiv\frac{\lambda N_{c}}{216\pi^{3}},\qquad K(z)\equiv1+z^{2}={U^{3}}/{U_{\kk}^{3}}.
\ee
To identify the four-dimensional meson fields, we make the
ansatz
\begin{align}
        &A_{\mu}(x^\mu,z)=\sum_{n=1}^{\infty}B_{\mu}^{\left(n\right)}(x^{\mu})\psi_{n}(z)\nonumber\\
&A_{z}(x^\mu,z)= \sum_{n=0}^{\infty}\varphi^{\left(n\right)}(x^{\mu})\phi_{n}(z) 
    \label{eq:separationAnsatz} 
\end{align}
for the five-dimensional gauge field using the complete sets $\left\{ \psi_{n}(z)\right\} _{n\geq1}$
and $\left\{ \phi_{n}(z)\right\} _{n\geq0}$ of normalizable functions
of $z$ with normalization conditions 
\be
\begin{split}
    &\kappa \int \d z K^{-1/3} \psi_{m}\psi_{n}=  \delta_{mn},\\
&\kappa  \int \d z K \phi_{m}\phi_{n}=\delta_{mn},
\end{split}\label{eq:normalizationConditions}
\ee
satisfying the completeness relations
\be
\begin{split}
    &\kappa\sum_{n}K^{-1/3}\psi_{n}(z)\psi_{n}(z^{\prime})=  \delta(z-z^{\prime}),\\
    &\kappa\sum_{n}K\phi_{n}(z)\phi_{n}(z')=\delta(z-z^{\prime}).\label{eq:completenessRelation}
\end{split}
\ee
With this ansatz, the fields $B_{\mu}^{\left(n\right)}$ and $\varphi^{\left(n\right)}$
have canonical kinetic terms; the eigenvalue equation 
\begin{eqnarray}
-K^{-1/3}\partial_{z}\left(K\partial_{z}\psi_{n}\right)= & \lambda_{n}\psi_{n},\label{eq:eomPsiN}
\end{eqnarray}
which can be used to relate the two complete sets via $\phi_{n}(z)\propto \partial_{z}\psi_{n}(z)$
for $\left(n\geq1\right)$, yields a mass term for $B_{\mu}^{\left(n\right)}$.
The remaining massless mode is given by $\phi_{0}(z)=1/\left(\sqrt{\pi\kappa}M_{\kk}K(z)\right)$.

Inserting the separation ansatz \eqref{eq:separationAnsatz} into the
DBI action \eqref{eq:5dDBI} and integrating over $z$, we obtain 
\begin{widetext}
\begin{eqnarray}
S_{\text{D8}}^{\text{DBI}}&=& -\Tr\int \d^4x \left[\left(\partial_{\mu}\varphi^{\left(0\right)}\right)^{2}+\sum_{n=1}^{\infty}\left(\frac{1}{2}\left(\partial_{\mu}B_{\nu}^{\left(n\right)}-\partial_{\nu}B_{\mu}^{\left(n\right)}\right)^{2}+
m_n^2\left(B_{\mu}^{\left(n\right)}-
m_n^{-1}\partial_{\mu}\varphi^{\left(n\right)}\right)^{2}\right)\right]\nonumber \\
 && +\left(\text{interaction terms}\right).
\end{eqnarray}    
\end{widetext}

The scalar fields $\varphi^{\left(n\right)}$ with $\left(n\geq1\right)$ can be absorbed by
the fields $B_{\mu}^{\left(n\right)}$, which
are interpreted as (axial) vector meson fields, with masses $m_n=\sqrt{\lambda_n}\MKK$ determined by the eigenvalue equation
for the normalizable modes \eqref{eq:eomPsiN}. 
The lightest vector mesons, identified with the rho and omega mesons, have $m_\rho=m_1=\sqrt{0.669314}\MKK$,
with the traditional value \cite{Sakai:2004cn,Sakai:2005yt} of $\MKK=949$ MeV corresponding to
$m_\rho=776.4$ MeV.

The remaining
field $\varphi^{\left(0\right)}$ is identified as the multiplet of massless pion fields produced by chiral symmetry breaking, which is realized geometrically by D8 and $\overline{\text{D8}}$-branes joining at $z=0$, with the $\mathrm{U}(N_f)$-valued Goldstone boson field given by the holonomy
\be\label{eq:Ux}
U(x)=e^{i\Pi^a(x)\lambda^a/f_\pi}=\mathrm P\,\exp i\int_{-\infty}^\infty \d z\, A_z(z,x),
\ee
where $\lambda^a=2T^a$ are Gell-Mann matrices including $\lambda^0=\sqrt{2/N_f}\mathbf{1}$. 
For $N_f=3$ we have 
\be\label{eq:Pix}
    \Pi(x)\equiv\Pi^a(x)T^a=\frac12\begin{pmatrix}
    \pi^0+\eta^8/\sqrt{3}+\eta^0\sqrt{2/3} & \sqrt{2} \pi^+ & \sqrt{2} K^+\\
    \sqrt{2} \pi^- & -\pi^0+\eta^8/\sqrt{3}+\eta^0\sqrt{2/3} & \sqrt{2} K^0 \\
    \sqrt{2} K^- & \sqrt{2} \bar K^0 & -2\eta^8/\sqrt{3}+\eta^0\sqrt{2/3}
    \end{pmatrix}.
\ee    

The pion decay constant is determined by
\be\label{fpi2}
f_\pi^2=\frac{\lambda N_c\MKK^2}{54\pi^4};
\ee
with the choice $f_\pi\approx 92.4$ MeV one obtains $\lambda\approx 16.63$. 
Following \BPR, we shall also consider the smaller value $\lambda\approx 12.55$ obtained by matching the large-$N_c$
lattice result for the string tension obtained in
Ref.~\cite{Bali:2013kia} (resulting in $f_\pi\approx 80.3$ MeV). A smaller 't Hooft coupling has also been
argued for in Ref.~\cite{Imoto:2010ef} from studies of the spectrum of higher-spin mesons in the WSS model. 
The downward
variation of $\lambda\approx 16.63\dots12.55$ will thus be
used as an estimate of the variability of the predictions of this model.

\subsection{Pseudoscalar masses}

In the WSS model, the U(1)$_A$ flavor symmetry is broken by an anomalous contribution of
order $1/N_c$ due to the $C_1$ Ramond-Ramond field, which gives rise to a 
Witten-Veneziano \cite{Witten:1979vv,Veneziano:1979ec} mass term for the singlet $\eta_0$ pseudoscalar
with \cite{Sakai:2004cn}
\be\label{mWV2}
m_{0}^2=\frac{2N_f}{f_\pi^2}\chi_g=\frac{N_f}{27\pi^2 N_c}\lambda^2\MKK^2,
\ee
where $\chi_g$ is the topological susceptibility.

For $N_f=N_c=3$, one has
$m_{0}=967\dots730$ MeV for $\lambda=16.63\dots12.55$, which
is indeed a phenomenologically interesting ballpark when
finite quark masses are added to the model by the
addition of an effective Lagrangian
\be\label{Lm}
\begin{split}
    &\mathcal{L}_m^{\mathcal M} \propto\Tr\left(\mathcal M\,U(x)+h.c.\right),\\
    &\mathcal M={\rm diag}(m_u,m_d,m_s).    
\end{split}
\ee
This deformation can be generated
by either worldsheet instantons \cite{Aharony:2008an,Hashimoto:2008sr}
or nonnormalizable modes of bifundamental fields corresponding to open-string tachyons \cite{0708.2839,Dhar:2008um,McNees:2008km,Niarchos:2010ki}.
Assuming for simplicity isospin symmetry, $m_u=m_d=\hat m$,
this leads to masses \cite{Brunner:2015oga}
    \be
m^2_{\eta,\eta'}=\frac12 m_0^2+m_K^2\mp\sqrt{\frac{m_0^4}{4}-\frac13 m_0^2(m_K^2-m_\pi^2)+(m_K^2-m_\pi^2)^2}
\ee
for the mass eigenstates
\bea
\eta &=& \eta_8 \cos\theta_P - \eta_0 \sin\theta_P\nn\\
\eta'&=& \eta_8 \sin\theta_P + \eta_0 \cos\theta_P,
\eea
with mixing angle
\be\label{thetaP}
\theta_P=\frac12\arctan\frac{2\sqrt2}{1-\frac32 {m_0^2}/({m_K^2-m_\pi^2})}.
\ee
Using
$m_\pi^2=m_{\pi_0}^2\approx (135 {\rm MeV})^2$ and
\be
m_K^2=\frac12(m_{K_\pm}^2+m_{K_0}^2)-
\frac12(m_{\pi_\pm}^2-m_{\pi_0}^2)\approx (495{\rm MeV})^2
\ee
as isospin symmetric parameters,
the WSS result $m_0\approx 967\dots730$ MeV for $\lambda=16.63\dots12.55$
leads to $\theta_P\approx -14^\circ\dots-24^\circ$
and 
$m_\eta\approx 520\dots 470$, $m_{\eta'}\approx 1080\dots 890$ MeV. 
In the following we shall consider this range of mixing angles in conjunction with the variation of $\lambda$, but we shall fix $m_\eta$ and $m_{\eta'}$
to their experimental values when evaluating phase space integrals.
In the radiative decay rates considered below, the explicit quark
masses will not modify the (chiral) results for the couplings; they only appear in phase space factors.

\subsection{Hadronic vector and axial vector meson decays}

Vertices for the hadronic decays of vector and axial vector meson
involving pseudoscalar mesons are contained in the second term
of the DBI action \eqref{eq:5dDBI}.
For the $\rho$ meson,
this contains the term (with indices restricted to the first two quark flavors)
\begin{align}
&\mathcal L_{\rho\pi\pi}=-g_{\rho\pi\pi} 
\varepsilon_{abc}(\partial_\mu \pi^a)
\rho^{b\mu}\pi^c,\nonumber\\
&g_{\rho\pi\pi}=
\int \d z \frac1{\pi K}\psi_1=33.98
\,\lambda^{-\frac12} N_c^{-\frac12},
\end{align}
yielding
$\Gamma_{\rho\to\pi\pi}=98.0\dots 130$ MeV for $\lambda=16.63\dots12.55$, which
somewhat underestimates the experimental result of $\approx 150$ MeV.

There is also a vertex involving one vector, one axial vector, and one pseudoscalar meson, which for the ground-state isotriplet mesons reads
\begin{align}\label{ga1rhopi}
&\mathcal L_{a_1\rho\pi}=g_{a_1\rho\pi} 
\varepsilon_{abc}a_\mu^a
\rho^{b\mu}\pi^c,\nonumber\\
&g_{a_1\rho\pi}=
2\MKK\sqrt{\frac{\kappa}{\pi}}\int\d z\psi_2'\psi_1={-34.43}\,\lambda^{-\frac12} N_c^{-\frac12}\MKK.
\end{align}

In the WSS model, the predicted mass of the $a_1$ meson, 1186.5 MeV, is rather close to the experimental result \cite{Workman:2022ynf} of 1230(40) MeV.
The predicted width for $a_1\to\rho\pi$ 
(already studied in \cite{Sakai:2005yt}) is 
$425\dots 563$ MeV, which is within the experimental result
for the total width of 250\dots 600 MeV (average value 420(35) MeV),
but according to \cite{CLEO:1999rzk} only 60\% of the three-pion decays
are due to S-wave $\rho\pi$ decays, whereas the latter saturate the
hadronic decays in the WSS model.

For the light quark flavors, these results for the decay rates of $\rho$ and $a_1$ seem to indicate that the WSS model is working quite well. When the mass of the
strange quark is included, a shortcoming of the model, which is shared
by many bottom-up holographic QCD models (see e.g.\ \cite{Abidin:2009aj}), 
is that the $\phi$ meson
remains degenerate with $\rho$ and $\omega$. In the following we shall
nevertheless also consider $K^*$ and $\phi$ mesons by simply raising
their masses in the resulting phase space factors while keeping their
vertices such as $g_{K^*K\pi}=g_{\phi KK}=g_{\rho\pi\pi}$ unchanged. 
The resulting widths, $\Gamma(K^*\to K\pi)=28\ldots 37$ MeV and
$\Gamma(\phi\to K\bar K)=2.12\ldots 2.82$ MeV, are between 40 and 20\% too small. These deviations are at least not dramatically larger than the one for the $\rho$ width, which amounts to 33\dots12\%; all appear to remain in the range to be expected for a large-$N$ approach. 

\section{Radiative Meson Decays}
\label{sec:rmd}

Before considering radiative decays of the experimentally elusive
glueballs, we shall
evaluate the predictions of the WSS model with nonzero quark masses for 
radiative decay widths of regular mesons
and compare with experimental data as far as available.
As discussed extensively in the second paper of Sakai and Sugimoto \cite{Sakai:2005yt},
holographic QCD models naturally provide a realization of vector meson
dominance (VMD) \cite{Gell-Mann:1961jim,Kroll:1967it,Sakurai:1969ss,Sakurai:1972wk} involving an infinite tower of vector mesons. There it was already
observed that the chiral WSS
model yields a result for $\Gamma(\omega\to\pi^0\gamma)$
which is roughly consistent with the experimental value. In the following
we shall recapitulate the results of \cite{Sakai:2005yt} and extend them
to the WSS model including quark masses and the Witten-Veneziano mass term.

\subsection{Vector meson dominance}

According to the holographic principle, 
non-normalizable modes are interpreted as external sources. 
This permits to study electromagnetic interactions to leading order by setting asymptotic values of the gauge field $A_\mu$ on the D8 branes according to \cite{Sakai:2005yt}
\be
\lim_{z\to\pm\infty}A_\mu(x,z)=A_{L,R\mu}(x)= eQA_{\mu}^{\text{em}}(x),
\ee
where $e$ is the electromagnetic coupling constant and $Q$ is the
electric charge matrix, given as
\begin{eqnarray}
Q & =\frac{1}{3}\left(\begin{array}{ccc}
2\\
 & -1\\
 &  & -1
\end{array}\right)
\end{eqnarray}
for the $N_{f}=3$ case. 
The ansatz \eqref{eq:separationAnsatz} changes to 
\be
\begin{split}
A_{\mu}(x^\mu,z)=&  A_{L\mu}(x^{\mu})\psi_{+}(z)+A_{R\mu}(x^{\mu})\psi_{-}(z)\\
&+\sum_{n=1}^{\infty}v_{\mu}^{n}(x^{\mu})\psi_{n}(z),    
\end{split}
\ee
with the functions $\psi_{\pm}(z)$ defined as 
\be
\psi_{\pm}(z)\equiv  \frac{1}{2}\left(1\pm\psi_{0}(z)\right),\qquad\psi_{0}(z)\equiv\frac{2}{\pi}\arctan z,
\ee
They satisfy \eqref{eq:eomPsiN} as non-normalizable zero modes,
because $\partial_{z}\psi_{\pm}(z)\propto\phi_{0}(z)\propto 1/K(z)$. 

To distinguish between vector and axial-vector fields we introduce the
notation 
\be
\begin{split}
    &\mathcal{V}_{\mu}\equiv\frac{1}{2}\left(A_{L\mu}+A_{R\mu}\right),\qquad  \mathcal{A}_{\mu}\equiv\frac{1}{2}\left(A_{L\mu}-A_{R\mu}\right),\\ &v_{\mu}^{n}\equiv B_{\mu}^{\left(2n-1\right)},\qquad a_{\mu}^{n}\equiv B_{\mu}^{\left(2n\right)},
\end{split}
\ee
so that
\be
\begin{split}
A_{\mu}(x^\mu,z)=&  \mathcal{V}_{\mu}(x^{\mu})+\mathcal{A}_{\mu}(x^{\mu})\psi_{0}(z)\\
&+\sum_{n=1}^{\infty}v_{\mu}^{n}(x^{\mu})\psi_{2n-1}(z)+\sum_{n=1}^{\infty}a_{\mu}^{n}(x^{\mu})\psi_{2n}(z).  
\end{split}\label{eq:modeExpansion}  
\ee

The first term in \eqref{eq:5dDBI} can then be expanded as 
\begin{eqnarray}
 \frac{\kappa}{2}\int\d z && \!\!\!\!\!\! K^{-1/3}F_{\mu\nu}^{2}\nonumber \\
&= & \frac{a_{\mathcal{V}\mathcal{V}}}{2}\text{tr}\left(\partial_{\mu}\mathcal{V}_{\nu}-\partial_{\nu}\mathcal{V}_{\mu}\right)^{2}\nonumber\\
&&+\frac{a_{\mathcal{A}\mathcal{A}}}{2}\text{tr}\left(\partial_{\mu}\mathcal{A}_{\nu}-\partial_{\nu}\mathcal{A}_{\mu}\right)^{2}\nonumber \\
 && +\frac{1}{2}\text{tr}\left(\partial_{\mu}v_{\nu}^{n}-\partial_{\nu}v_{\mu}^{n}\right)^{2}+\frac{1}{2}\text{tr}\left(\partial_{\mu}a_{\nu}^{n}-\partial_{\nu}a_{\mu}^{n}\right)^{2}\nonumber \\
 && +a_{\mathcal{V}v^{n}}\text{tr}\left(\left(\partial_{\mu}\mathcal{V}_{\nu}-\partial_{\nu}\mathcal{V}_{\mu}\right)\left(\partial_{\mu}v_{\nu}^{n}-\partial_{\nu}v_{\mu}^{n}\right)\right)\nonumber\\
 &&+a_{\mathcal{A}a^{n}}\left(\left(\partial_{\mu}\mathcal{A}_{\nu}-\partial_{\nu}\mathcal{A}_{\mu}\right)\left(\partial_{\mu}a_{\nu}^{n}-\partial_{\nu}a_{\mu}^{n}\right)\right)\nonumber \\
 && +\left(\text{interaction terms}\right),\label{eq:SkinMix}
\end{eqnarray}
with coupling constants 
\begin{equation}
\begin{split}
&a_{\mathcal{V}v^{n}}= \kappa \int dz K^{-1/3}\psi_{2n-1},\quad a_{\mathcal{V}\mathcal{V}}=\kappa \int dz K^{-1/3},\\
&a_{\mathcal{A} a^{n}}= \kappa \int dz K^{-1/3}\psi_{2n}\psi_{0},\quad a_{\mathcal{A} A}=\kappa \int dz K^{-1/3}\psi_{0}^{2}    ,
\end{split}
\label{eq:mixingTerms}
\end{equation}
mixing the photon field $\mathcal{V}$ with every vector meson $v^{n}$. The coefficients $a_{\mathcal{V}\mathcal{V}}$ and $a_{\mathcal{A}\mathcal{A}}$ are divergent, since the external
fields correspond to non-normalizable modes in the radial direction, and need to be renormalized to canonical values.
The photon field $\mathcal{V}$ does not appear in the interaction terms of
this model and can only couple via the mixing \eqref{eq:mixingTerms}, fully
realizing VMD. 
Alternatively, it is possible to
perform a field redefinition to diagonalize the action and to get rid of the
mixing terms, thus producing new interaction terms coupling mesons
to photons.

\subsection{Radiative decays of pseudoscalars and vector mesons}
\label{sec:raddecpscvm}

The relevant vertices for radiative decays of pseudoscalars and (axial) vector mesons come from the Chern-Simons term
\begin{align}
    S_{CS}\supset &\, T_{8}\int\tr\left(\exp\left(2\pi\alpha'F_{2}+B_{2}\right)\wedge C_{3}\right)\nonumber \\
    \supset\, & \frac{N_{c}}{96\pi^{2}}\epsilon^{\mu\nu\rho\sigma z}\int\tr\left(3A_{z}F_{\mu\nu}F_{\rho\sigma}-4A_{\mu}\partial_{z}A_{\nu}F_{\rho\sigma}\right),\label{eq:5dCS}
\end{align}
where we have used partial integration. 

Inserting the mode expansion
\eqref{eq:modeExpansion} and integrating over the radial coordinate
we obtain for the interaction term involving two vectors and one pseudoscalar
\begin{align}
    \mathcal{L}_{\Pi v^{m}v^{n}}= & \frac{N_{c}}{4\pi^{2}f_{\pi}}c_{v^{n}v^{m}}\epsilon^{\mu\nu\rho\sigma}\tr\left(\Pi \partial_{\mu} v_{\nu}^{n}\partial_{\rho} v_{\sigma}^{m}\right),
\end{align}
with coupling constants
\begin{equation}
        c_{v^{n}v^{m}}=\frac{1}{\pi}\int\d zK^{-1}\psi_{2n-1}\psi_{2m-1} =
        \left\{\frac{1350.83}{\lambda\Nc},\ldots\right\}
\end{equation}
as studied in \cite{Sakai:2005yt}, where numerical results for the coefficients beyond $c_{v^1v^1}$ given above can be found.

\subsubsection{Vector meson \texorpdfstring{$1\gamma$-decays}{1 photon}}

Using VMD, we can calculate the interaction term for the radiative decay of a vector meson into a pseudoscalar and one photon as
\begin{equation}
    \mathcal{L}_{\Pi\V v^{n}}  =\frac{N_{c}}{4\pi^{2}f_{\pi}}c_{\V v^{n}}\epsilon^{\mu\nu\rho\sigma}\tr\left(\pion\partial_{\mu} v_{\nu}^{n}\partial_{\rho}\V_{\sigma}+\pion\partial_{\mu}\V_{\nu}\partial_{\rho} v_{\sigma}^{n}\right),
\end{equation}
with coupling 
\begin{equation}
\begin{split}
        c_{\V v^{n}}= & \sum_{m}c_{v^{n} v^{m}}a_{\V v^{m}}=\frac{1}{\pi}\int\d zK^{-1}\psi_{2n-1}\\
        &=\{33.9839,\dots\}(\Nc\lambda)^{-1/2},
\end{split}
\end{equation}
where we have used the completeness relation \eqref{eq:completenessRelation}
to eliminate the summed-over modes.

Performing the polarization sums we get 
\begin{equation}
    \begin{split}
        &\left|\mathcal{M}_{\left(v^{n}\rightarrow\pion\V\right)}\right|^{2}= \sum_{\left(v^{n}\right)}\sum_{\left(\V\right)}\frac{1}{3}\epsilon_{\mu}^{\left(v^{n}\right)}\epsilon_{\nu}^{\left(v^{n}\right)*}\epsilon_{\rho}^{\left(\V\right)}\epsilon_{\sigma}^{\left(\V\right)*}\mathcal{M}_{\left(\pion v^{n}\V\right)}^{\mu\rho}\mathcal{M}_{\left(\pion v^{n}\V\right)}^{\nu\sigma*}\nonumber \\
    &= \frac{c_{\V v^{n}}^{2}e^{2}\Nc^{2}}{96\pi^{4}f_{\pi}^{2}}\left(\tr\left(T_{\pion}T_{v^{n}}Q\right)+\tr\left(T_{\pion}QT_{v^{n}}\right)\right)^{2}\left(m_{\pion}^{2}-m_{v^{n}}^{2}\right)^{2}.
    \end{split}
\end{equation}
The partial width then reads 
\begin{align}
    \Gamma_{v^{n}\rightarrow\pion\gamma}= & \frac{1}{8\pi}\left|\mathcal{M}_{\left(v^{n}\rightarrow\pion\V\right)}\right|^{2}\frac{\left|\mathbf{p}_v\right|}{m_v^{2}}.
\end{align}

\begin{table}
    \centering{}\bigskip{}
    \begin{tabular}{lcc}
    \toprule
     & $\Gamma^{{\rm exp.}}${[}keV{]} & $\Gamma^{{\rm WSS}}${[}keV{]}  \tabularnewline
    \colrule
    $\pi^0\rightarrow2\gamma$  & 
    0.00780(12) 
    & 0.00773\dots0.0102  \tabularnewline
    $\eta\rightarrow2\gamma$ & 0.515(18) & 0.480\dots0.978\tabularnewline
    $\eta^{\prime}\rightarrow2\gamma$ & 4.34(14)  & 5.72\dots5.87\dots5.75 \tabularnewline
\hline
    $\rho^{0}\to\pi^{0}\gamma$ & 70(12) & 56.2\dots98.6 \tabularnewline
    $\rho^{\pm}\to\pi^{\pm}\gamma$ & 68(7) &  56.2\dots98.6 \tabularnewline
    $\rho^0\to\eta\gamma$ & 45(3) & 40.3\dots{90.5}\tabularnewline
    $\omega\to\pi^0\gamma$ & 725(34) & $521$\dots 915 \tabularnewline
    $\omega\to\eta\gamma$ & 3.9(4) & 4.87\dots10.9\tabularnewline
    $\eta'\to\rho^0\gamma$  & 55.4(1.9)$^\mathrm{fit}$,68(7)$^\mathrm{av.}$ & 54.1\dots59.2\dots58.5 \tabularnewline
    $\eta'\to\omega\gamma$ & 4.74(20)$^\mathrm{fit}$,5.8(7)$^\mathrm{av.}$ & 5.37\dots5.89\dots5.81\tabularnewline
    $\phi\to\pi^0\gamma$   & 5.6(2)  & 0 \tabularnewline
    $\phi\to\eta\gamma$  & 55.3(1.2) & 84.7\dots92.8\dots91.6\tabularnewline
    $\phi\to\eta'\gamma$   & 0.264(10) & 0.525\dots 1.18\tabularnewline
    $K^{*0}\to K^0\gamma$& 116(10) & 124\dots218 \tabularnewline
    $K^{*\pm}\to K^\pm\gamma$ & 50(5) & 31.0\dots54.5 \tabularnewline
    \botrule
    \end{tabular}
    \caption{Results for various radiative decay widths of pseudoscalar and vector mesons involving vector and pseudoscalar
    mesons, with 't Hooft coupling $\lambda=16.63\dots12.55$ ($\lambda=16.63$ is the traditional \cite{Sakai:2004cn,Sakai:2005yt} value matching $f_\pi=92.4$ MeV; $\lambda=12.55$ an alternative choice matching the large-$N$ string tension at the expense of $f_\pi$). For nonmonotonic dependence on $\lambda$
    intermediate extremal values are also given. Ideal mixing is assumed for $\omega$ and $\phi$, {fixing the WSS result for $\phi\to\pi^0\gamma$ to zero}.
    Experimental results are from the PDG \cite{Workman:2022ynf} except for the $\pi^0$ width, which is from \cite{PrimEx-II:2020jwd}.
    \label{tab:vectorDecays}}
\end{table}

\subsubsection{Pseudoscalar meson \texorpdfstring{$2\gamma$-decays}{2 photon} }

Employing VMD a second time, we can derive the interaction term for a decay
of a pseudoscalar meson in two photons
\begin{align}
    \mathcal{L}_{\Pi\V\V}= & -\frac{N_{c}}{4\pi^{2}f_{\pi}}c_{\V\V}\epsilon^{\mu\nu\rho\sigma}\tr\left(\pion\partial_{\mu}\V_{\nu}\partial_{\rho}\V_{\sigma}\right),
\end{align}
where the sum over the entire tower of vector mesons yields
\begin{align}
    c_{\V\V}= & \sum_{m}c_{\V v^{m}}a_{\V v^{m}}=\frac{1}{\pi}\int\d zK^{-1}=1,
    \end{align}
leading to the standard result
\begin{align}
    \Gamma_{\Pi\rightarrow\gamma\gamma}= & \frac{1}{8\pi}\left|\mathcal{M}_{\left(\Pi\rightarrow\V\V\right)}\right|^{2}\frac{\left|\mathbf{p}_{\gamma}\right|}{m_{\Pi}^{2}}\frac{1}{2}
\end{align}
with
\begin{align}\label{MPiVV}
    \left|\mathcal{M}_{\left(\Pi\rightarrow\V\V\right)}\right|^{2}= & \frac{e^{4}\Nc^{2}}{4\pi^{4}f_{\pi}^{2}}\left(\tr\left(T_{\pion}Q^{2}\right)\right)^{2}m_{\Pi}^{4}.
\end{align}

The numeric results for the various radiative decays involving one pseudoscalar and two vector particles are summarized in Table \ref{tab:vectorDecays}
for $\lambda=16.63\dots12.55$. As mentioned above, $\lambda=16.63$ is the traditional \cite{Sakai:2004cn,Sakai:2005yt} value matching $f_\pi=92.4$ MeV,
whereas $\lambda=12.55$ is an alternative choice matching the large-$N$ string tension at the expense of $f_\pi$. 
The decay rate for $\pi^0$ is therefore close to the experimental value only
for the first value of $\lambda$, but the partial widths of the decays $\rho$ and $\omega$ into $\pi\gamma$ are reproduced by an intermediate value of $\lambda$.

In processes involving $\eta$ and $\eta'$, we have used the
pseudoscalar mixing angle following from (\ref{thetaP}), which varies as 
$\theta_{P}\approx-14^{\circ}\dots-24^{\circ}$ when $\lambda=16.63\dots12.55$,
{which enters the flavor matrix $T_\Pi$ in \eqref{MPiVV}}.
Here the dependence on $\lambda$ is nonmonotonic, 
{because also $f_\pi$ in the prefactor depends on $\lambda$};
Table \ref{tab:vectorDecays} also gives the extremal values attained
at intermediate values of $\lambda$. 

The vector couplings in the WSS model {augmented by quark masses according to \eqref{Lm}} are flavor-symmetric, but we distinguish $\phi$ and $\omega$ mesons through their experimental masses.
{The undetermined mixing of $\phi$ and $\omega$ could be fixed by fitting for example the small ratio of the widths for their decays into $\pi^0\gamma$, 5.6/725, which yields a mixing angle close to ideal mixing, $\theta_V=\theta_V^\mathrm{ideal}+3.32^\circ$, as in \cite{Ambrosino:2009sc}. However, here and in the following we shall assume completely}
ideal mixing for simplicity, {which eliminates $\phi\to\pi\gamma$ but does not change the other partial widths of $\phi$ significantly.}
This gives generally good results for decays involving $\omega$, but larger discrepancies with experiment for $\phi$ mesons {irrespective of the precise value of $\theta_V$}.\footnote{With a $\phi$-$\omega$ mixing angle of $3.32^\circ$ above ideal mixing \cite{Ambrosino:2009sc}, we would have $\Gamma^\mathrm{WSS}(\phi\to\pi\gamma)=4\ldots7$ keV, consistent with experiment; the result for $\omega\to\eta\gamma$ would be somewhat closer to the experimental value, but the one for $\eta'\to\omega\gamma$ further off.} 
{Note that the standard value of $M_\mathrm{KK}=949$ MeV, which we are using, is chosen such that the $\rho$ mass is reproduced, which is rather close to the mass of the $\omega$ meson, but less suitable for the $\phi$ meson.}

\subsection{Radiative axial-vector decays}

From the 5-dimensional CS-term \eqref{eq:5dCS} we can also extract
a term including two vector mesons and one axial-vector meson
\begin{align}
    \mathcal{L}_{v^{m}v^{n}a^{p}}= & -\frac{N_{c}}{4\pi^{2}}d_{v^{m}v^{n}a^{p}}\epsilon^{\mu\nu\rho\sigma}\tr\left(v_{\mu}^{m}a_{\nu}^{p}\partial_{\rho} v_{\sigma}^{n}\right),
\end{align}
with
\begin{align}
    d_{v^{m}v^{n}a^{p}}= & \int\d z\psi_{2m-1}\psi_{2n-1}^{\prime}\psi_{2p},\label{eq:axialCoupling}
\end{align}
where we again made use of partial integration.

As noted already in \cite{Sakai:2005yt} and observed before in other holographic models \cite{Son:2003et,DaRold:2005mxj,Hirn:2005nr} 
as well as in the hidden local symmetry approach of \cite{Bando:1984ej}, the vertex for the decay of an axial vector meson into a pseudoscalar and a photon, which would have to come from the DBI part of the action, vanishes,\footnote{In the hidden local symmetry approach, $a_1\to\pi\gamma$ has been included by adding higher-derivative terms
to the action \cite{Bando:1987ym}.}
even though there is a nonvanishing vertex for $a_1^\pm\to\pi^\pm\rho^0$, see
\eqref{ga1rhopi}. But the corresponding coupling for an on-shell photon is obtained by replacing $\psi_1$ therein by a unity, leading to
\be
g_{a_1\pi\mathcal{V}}=
2\MKK\sqrt{\frac{\kappa}{\pi}}\int\d z\psi_2'=0,
\ee
implying a cancellation between the contribution from the lowest vector meson and the remaining tower.
Indeed, the experimental result for $a_1^\pm\to\pi^\pm\gamma$ is much smaller than
expected from naive VMD \cite{Zielinski:1984au}.

\subsubsection{Axial-vector \texorpdfstring{$1\gamma$-decays}{1 photon} }

Employing VMD once we obtain for the interaction between one axial vector meson, one vector meson and one photon
\begin{align}
    \mathcal{L}_{\V v^{n}a^{p}}= & -\frac{N_{c}}{4\pi^{2}}d_{\V v^{n}a^{p}}\epsilon^{\mu\nu\rho\sigma}\tr\left( v_{\mu}^{n}a_{\nu}^{p}\partial_{\rho}\V_{\sigma}\right),
\end{align}
with
\begin{align}
    d_{\V v^{n}a^{p}}= & \int\d z\psi_{2n-1}^{\prime}\psi_{2p}=\{-2497.14,\dots\}N_{c}^{-1}\lambda^{-1},
\end{align}
where we had to sum over the radial mode without the derivative to
get a non-vanishing result since the bulk-to-boundary propagator associated
to an on-shell photon is constant. The amplitudes for the decay $a\rightarrow v\V$,
for the different combinations of polarizations read
\begin{align}
    \left|\mathcal{M}_{-101}^{a^{p}\rightarrow v^{n}\V}\right|= & \frac{d_{\V v^{n}a^{p}}\left(m_{a^{p}}^{2}-m_{v^{n}}^{2}\right)N_{c}}{8m_{v^{n}}\pi^{2}}\tr\left(eQT_{a^{p}}T_{v^{n}}\right)\nonumber \\
    \left|\mathcal{M}_{-110}^{a^{p}\rightarrow v^{n}\V}\right|= & \frac{d_{\V v^{n}a^{p}}\left(m_{a^{p}}^{2}-m_{v^{n}}^{2}\right)N_{c}}{8m_{a^{p}}\pi^{2}}\tr\left(eQT_{a^{p}}T_{v^{n}}\right),
\end{align}
which yields
\begin{align}
    \left|\mathcal{M}_{a^{p}\rightarrow v^{n}\V}\right|^{2}= & \frac{1}{3}\left(2\left|\mathcal{M}_{-101}^{a^{p}\rightarrow v^{n}\V}\right|^{2}+2\left|\mathcal{M}_{-110}^{a^{p}\rightarrow v^{n}\V}\right|^{2}\right)\nonumber \\
    = & \frac{d_{\V v^{n}a^{p}}^{2}\left(m_{a^{p}}^{2}-m_{v^{n}}^{2}\right)^{2}\left(m_{a^{p}}^{2}+m_{v^{n}}^{2}\right)N_{c}^{2}}{96\pi^{4}m_{a^{p}}^{2}m_{v^{n}}^{2}}\left(\tr\left(eQT_{a^{p}}T_{v^{n}}\right)\right)^{2}.
\end{align}

The decay width is given by
\begin{align}
    \Gamma_{a^{p}\to v^{n}\gamma}= & \frac{1}{8\pi}\left|\mathcal{M}_{a^{p}\rightarrow v^{n}\V}\right|^{2}\frac{\left|\mathbf{p}_{\V}\right|}{m_{a^{p}}^{2}},
\end{align}
and the numerical results are listed in Table \ref{tab:radAxialDecay}.

\begin{table}
    \centering{}\bigskip{}
    \begin{tabular}{lcc}
    \toprule
    &  $\Gamma^{\text{exp}}${[}keV{]}  & $\Gamma^{\text{WSS}}${[}keV{]}\tabularnewline \colrule
    $a_{1}(1260)\rightarrow\rho\gamma$   &  & 28.9\dots50.8\tabularnewline 
    $a_{1}(1260)\rightarrow\omega\gamma$  &  & 247\dots434\tabularnewline
    $f_{1}(1285)\rightarrow\rho\gamma$   & 1380(300)\dots640(240) 
    & 295\dots518$|$270\dots 473 \tabularnewline 
    $f_{1}(1285)\rightarrow\omega\gamma$  &  & 31.3\dots54.9$|$28.6\dots50.2\tabularnewline
    $f_{1}(1285)\rightarrow\phi\gamma$   & 17(7)  & 2.44\dots4.29$|$3.97\dots 6.98\tabularnewline
    $f_{1}(1420)\rightarrow\rho\gamma$   &  & 73.0\dots128$|$119\dots209\tabularnewline 
    $f_{1}(1420)\rightarrow\omega\gamma$   &  & 7.80\dots13.7$|$12.7\dots22.3\tabularnewline
    $f_{1}(1420)\rightarrow\phi\gamma$   & 164(55)  & 52.9\dots92.9$|$48.3\dots84.8\tabularnewline\botrule 
    \end{tabular}\caption{Radiative axial-vector meson decay with $\lambda=16.63\dots12.55$ and two values of the $f_1$ mixing angle
    $\theta_{f}=20.4^{\circ}|26.4^\circ$. 
    Experimental values are from the PDG \cite{Workman:2022ynf} with the exception of the lower values for $f_1(1285)\to\rho\gamma$, which are from VES \cite{Amelin:1994ii}; Zanke et al.~\cite{Zanke:2021wiq} propose here as experimental average 950(280) keV.
    }
    \label{tab:radAxialDecay} 
\end{table}

The PDG \cite{Workman:2022ynf} gives experimental results only for
the $f_1$ mesons, which in the WSS model have the same mass as the $a_1$ meson.
Besides extrapolating to their experimental masses we consider also two possible values (motivated below) for the mixing angle for the $f_1$ and $f_1'$ mesons
using the 
convention
\begin{align}\label{f1mixing}
    |f_1(1285) \rangle&=\cos \theta_f|\bar n n\rangle-\sin \theta_f|\bar s s\rangle \nonumber\\
    |f_1(1420) \rangle&=\sin \theta_f |\bar n n\rangle +\cos \theta_f |\bar s s\rangle 
\end{align}
so that ideal mixing corresponds to $\theta_f=0$.

{In Table \ref{tab:radAxialDecay}, the $\phi$-$\omega$ mixing is again assumed to be ideal. A value a bit above ideal mixing increases somewhat the branching ratio of $\phi\gamma$ over $\omega\gamma$ for $f_1(1285)$, while decreasing it for $f_1(1420)$.}

\subsubsection{Axial-vector \texorpdfstring{$2\gamma$-decays}{2 photon} }

As mentioned above, the radial derivative of
the bulk-to-boundary propagator for a photon vanishes for on-shell photons,
which implies that in accordance with the Landau-Yang theorem
at least one photon in the two-photon-decay of an axial vector meson has to be off-shell.
Denoting the virtual photon by $v^{*}$ we have
\begin{align}
    d_{\V v^{*}a^{p}}= & \int\d z\mathcal{J}^{\prime}\psi_{a^{p}},
\end{align}
where we have introduced the (off-shell) bulk-to-boundary propagator
$\mathcal{J}$ defined by
\begin{align}\label{JQz}
    \left(1+z^{2}\right)^{1/3}\partial_{z}\left[\left(1+z^{2}\right)\partial_{z}\mathcal{J}\right] & =\frac{Q^{2}}{M_{\kk}^{2}}\mathcal{J}.
\end{align}
Since we are only interested in the low $Q$ regime we make the ansatz
\begin{align}
    \mathcal{J}(Q,z) &=  1+\frac{Q^{2}}{M_{\kk}^{2}}\alpha(z)+\mathcal{O}(Q^{4})
\end{align}
satisfying 
\begin{align}
    \left(1+z^{2}\right)^{1/3}\partial_{z}\left[\left(1+z^{2}\right)\partial_{z}\alpha\right] & =1.
\end{align}
With the solution 
\begin{align}
    \partial_{z}\alpha= & \frac{z}{\left(1+z^{2}\right)}\;{}_{2}F_{1}\left(\frac{1}{3},\frac{1}{2},\frac{3}{2},-z^{2}\right)
\end{align}
we obtain for the relevant coupling constant 
\begin{align}
    d_{\V v^{*}a^{p}}= & \frac{Q^{2}}{M_{\kk}^{2}}\int\d z\alpha^{\prime}\psi_{a^{p}}+\mathcal{O}(Q^{4})\nonumber \\
    = & \frac{Q^{2}}{M_{\kk}^{2}}c_{\V v^{*}a^{p}}+\mathcal{O}(Q^{4})
\end{align}
with
\begin{align}
    c_{\V v^{*}a^{p}}= & 101.309N_{c}^{-1/2}\lambda^{-1/2}.
\end{align}
\begin{widetext}
The decay widths then read 
\begin{align}
    \Gamma(f(1285)\rightarrow\gamma_{L}^{*}\gamma_{T})= & \frac{2}{3}\left(\frac{c_{\V v^{*}a}m_{a}^{2}N_{c}}{8\pi^{2}M_{\kk}^{2}}\right)^{2} \frac{1}{8\pi} \frac{|\mathbf{p}|}{m_{a}^{2}} \left(\frac{5e^{2}}{18} \cos\theta_{f} - \frac{e^{2}}{9\sqrt{2}} \sin\theta_{f}\right)^{2}Q^{2}+\mathcal{O}(Q^{4})\nonumber \\
    \Gamma(f(1285)\rightarrow\gamma_{T}^{*}\gamma_{T})= & \mathcal{O}\left(Q^{6}\right).
\end{align}
and
\begin{align}
    \Gamma(f(1420)\rightarrow\gamma_{L}^{*}\gamma_{T})= & \frac{2}{3}\left(\frac{c_{\V v^{*}a}m_{a}^{2}N_{c}}{8\pi^{2}M_{\kk}^{2}}\right)^{2}\frac{1}{8\pi}\frac{|\mathbf{p}|}{m_{a}^{2}} \left( \frac{5e^{2}}{18}\sin\theta_{f}+\frac{e^{2}}{9\sqrt{2}}\cos\theta_{f}\right)^{2}Q^{2}+\mathcal{O}(Q^{4}).
\end{align}    
\end{widetext}

In the literature one usually finds the values for the so-called equivalent
photon rate
\begin{align}
    \tilde{\Gamma}_{\gamma\gamma}= & \lim_{Q^{2}\rightarrow0}\frac{m_{a}^{2}}{Q^{2}}\frac{1}{2}\Gamma\left(a\rightarrow\gamma_{L}^{*}\gamma_{T}\right),
\end{align}
which are listed in \ref{tab:axialTFF}.

\begin{table}
    \centering{}\bigskip{}
    \begin{tabular}{lcc}
    \toprule
    & $\tilde{\Gamma}_{\gamma\gamma}^{\text{exp}}${[}keV{]}  & $\tilde{\Gamma}_{\gamma\gamma}^{\text{WSS}}${[}keV{]}\tabularnewline\colrule 
    $a_1\left(1260\right)$ & - & 1.60\dots2.12$|$1.39\dots1.85 \tabularnewline
    \colrule $f_1\left(1285\right)$   & 3.5(8)  & 3.84\dots5.09$|$2.39\dots3.17 \tabularnewline
    $f_1\left(1420\right)$ & 3.2(9)  & 3.50\dots4.64$|$2.19\dots2.90\tabularnewline\botrule
    \end{tabular}
    \caption{Equivalent photon rates of axial vector mesons for two values of the $f_1$ mixing angle
    $\theta_{f}=20.4^{\circ}|26.4^\circ$
    (in the latter case with $M_\mathrm{KK}$ rescaled such that $m_a$ is raised to the experimental value which reduces $\xi$ in (\ref{thetaf-fit}) to zero); the range denoted by dots corresponds again to $\lambda=16.63\dots 12.55$, where
    only the first value is matching the axial anomaly exactly.
     Experimental values from L3 \cite{Achard:2001uu,Achard:2007hm}, see also \cite{Zanke:2021wiq}.
     }
    \label{tab:axialTFF} 
\end{table}

The mixing angle is inferred from
\begin{equation}\label{thetaf-fit}
    \tan^2\left(\theta_f-\arctan\frac{\sqrt{2}}{5}\right)=\left(\frac{m_{f_1}}{m_{f_1'}}\right)^{1+\xi} \frac{\tilde{\Gamma}_{\gamma\gamma}^{f_1'\text{exp}}}{\tilde{\Gamma}_{\gamma\gamma}^{f_1\text{exp}}},
\end{equation}
where the usual assumption of $\xi=0$ leads to $\theta_f=26.4^\circ$,
corresponding to the central value of $\theta_A=62(5)^\circ$ in \cite{Zanke:2021wiq}.
However, in the WSS the coupling $d_{\V v^{*}a^{p}}$ is proportional to $1/M_{\kk}^{2}$, which leads to a scaling of
$\tilde \Gamma_{\gamma\gamma}$ with four additional powers of $m_a$, i.e.\ $\xi=4$, resulting in $\theta_f=20.4^\circ$.

In Tables \ref{tab:radAxialDecay} and \ref{tab:axialTFF} we
consider two possible extrapolations to axial vector mesons
with realistic masses. In the first we keep the parameters
of the theory unchanged in the expressions for the couplings and use
the measured masses only in kinematical factors,
which leads to $\xi=4$ and $\theta_f=20.4^\circ$; 
in the second we rescale $M_{\kk}$ proportional to
$m_a^{\mathrm{exp}}/m_a^{\mathrm{WSS}}$ such that $\xi=0$ and
$\theta_f=26.4^\circ$.

While the predictions for the equivalent photon rate for
the $f_1$ mesons (shown in Table \ref{tab:axialTFF})
agree well with the experimental result for the standard choice
of $\lambda=16.63$ and $\theta_f=20.4^\circ$,
the 1-$\gamma$ decay rates are significantly underestimated.
In contrast to the radiative decays of vector mesons, lowering
$\lambda$ does not increase the rates sufficiently to cover the experimental
results. Unfortunately no experimental results are available for
isotriplet axial vector mesons, where the WSS model is generally
performing best.

\section{Glueballs in the Witten-Sakai-Sugimoto model}
\label{sec:GBs}

Glueballs are realized in the WSS model as fluctuations of the background
in which the probe D8 branes are placed, where certain superselection rules are applied. In particular states with odd parity in the extra circle along $\tau$ are discarded, as well as Kaluza-Klein modes of the compact $S^1$ and $S^4$ subspaces. The resulting glueball spectrum was
discussed in \cite{Brower:2000rp}, where the lift of \eqref{eq:background}
to 11-dimensional supergravity is used. 
In the following we shall consider
scalar, tensor, and pseudoscalar 
glueballs,
for which hadronic decays have been worked out
in the WSS model in \cite{Hashimoto:2007ze,Brunner:2015oqa,Brunner:2015yha,Brunner:2015oga,Leutgeb:2019lqu} 
and which we review and update in Appendix \ref{app:hadrGBdecays} in some detail for the scalar and tensor glueballs.

The lift of a type IIA string-frame metric to 11-dimensional supergravity
is given by the relation 
\begin{equation}
    \begin{split}
            \d s^{2}&= G_{MN}\d x^{M}\d x^{N}\\
            &=e^{-2\phi/3}g_{AB}\d x^{A}\d x^{B}+e^{4\phi/3}\left(\d x^{11}+A_{B}\d x^{B}\right)^{2},
    \end{split}
\end{equation}
with $M,N=0,\dots10$ and $A,B=0,\dots9$, omitting the 11th index. By
introducing the radial coordinate $r$ related to $U$ by $U=\frac{r^{2}}{2L}$,
we get the lifted metric
    \begin{equation}
    \d s^{2}=\frac{r^{2}}{L^{2}}\left[f(r)\d x_{4}^{2}+\eta_{\mu\nu}\d x^{\mu}\d x^{\nu}+\d x_{11}^{2}\right]+\frac{L^{2}}{r^{2}}\frac{\d r^{2}}{f\left(r\right)}+\frac{L^{2}}{4}\d\Omega_{4}^{2},
\end{equation}    
and the field strength $F_{\alpha\beta\gamma\delta}=\frac{6}{L}\sqrt{g_{S^{4}}}\epsilon_{\alpha\beta\gamma\delta}$,
which are solutions to the equations of motion following from the 
unique supergravity action
\begin{equation}
    2\kappa_{11}^{2}S_{11}=\int\mathrm{d}^{11}x\sqrt{-G}\left(R-\frac{1}{2}|F_4|^2\right)-\frac{1}{3!}\int A_3\wedge F_4\wedge F_4.
\end{equation}

Scalar and tensor glueball modes appear as normalizable modes of metric fluctuations $\delta G$, which translate to perturbations
of the type-IIA string metric and dilaton through
\begin{align}
    g_{\mu\nu}= & \frac{r^{3}}{L^{3}}\left[\left(1+\frac{L^{2}}{2r^{2}}\delta G_{11,11}\right)\eta_{\mu\nu}+\frac{L^{2}}{r^{2}}\delta G_{\mu\nu}\right]\nonumber \\
    g_{44}= & \frac{r^{3}f}{L^{3}}\left[1+\frac{L^{2}}{2r^{2}}\delta G_{11,11}+\frac{L^{2}}{r^{2}f}\delta G_{44}\right]\nonumber \\
    g_{rr}= & \frac{L}{rf}\left[1+\frac{L^{2}}{2r^{2}}\delta G_{11,11}+\frac{r^{2}f}{L^{2}}\delta G_{rr}\right]\nonumber \\
    g_{r\mu}= & \frac{r}{L}\delta G_{r\mu}\nonumber \\
    g_{\Omega\Omega}= & \frac{r}{L}\left(\frac{L}{2}\right)^{2}\left(1+\frac{L^{2}}{2r^{2}}\delta G_{11,11}\right)\nonumber \\
    e^{4\phi/3}= & \frac{r^{2}}{L^{2}}\left(1+\frac{L^{2}}{r^{2}}\delta G_{11,11}\right).
\end{align}
Inducing these metric fluctuations to the world volume of the D8-brane
system described by the action \eqref{eq:9dAction}, \cite{Hashimoto:2007ze} calculated interaction vertices of the lightest scalar glueball with mesons,
which was revisited and extended in \cite{Brunner:2015oqa}. 

Pseudoscalar, vector, and pseudovector glueballs appear as fluctuations of
the type-IIA 
form fields; glueballs with higher spin
would need a stringy description beyond the supergravity approximation \cite{Sonnenschein:2018fph}.

\subsection{Exotic and dilatonic scalar glueballs}

Superficially, the emerging glueball spectrum resembles the one found in  lattice calculations (see Fig.~1 in \cite{Brunner:2015oga}),
containing a lightest scalar glueball with a mass below that of the tensor glueball,
whereas most other holographic models have the scalar glueball degenerate with the tensor. This is achieved by an ``exotic'' polarization of the bulk metric involving the extra compact dimension $(\tau)$ separating the D8-branes,
\begin{align}\label{GEmetricfluctuations}
    \delta G_{\tau\tau} & =-\frac{r^{2}}{\mathcal{N}_{E}\,L^{2}}f(r)S_{4}(r)G_{E}(x^{\sigma}),\nonumber \\
    \delta G_{\mu\nu} & =\frac{r^{2}}{\mathcal{N}_{E}\,L^{2}}S_{4}(r)\left[\frac{1}{4}\eta_{\mu\nu}-\left(\frac{1}{4}+\frac{3r_{\kk}^{6}}{5r^{6}-2r_{\kk}^{6}}\right)\frac{\partial_{\mu}\partial_{\nu}}{M^{2}}\right]G_{E}(x^{\sigma}),\nonumber \\
    \delta G_{11,11} & =\frac{r^{2}}{\mathcal{N}_{E}\,4L^{2}}S_{4}(r)G_{E}(x^{\sigma}),\nonumber \\
    \delta G_{rr} & =-\frac{L^{2}}{\mathcal{N}_{E}\,r^{2}f(r)}\frac{3r_{\kk}^{6}}{5r^{6}-2r_{\kk}^{6}}S_{4}(r)G_{E}(x^{\sigma}),\nonumber \\
    \delta G_{r\mu}=\delta G_{\mu r} & =\frac{90r^{7}r_{\kk}^{6}}{\mathcal{N}_{E}\,M^{2}L^{2}\left(5r^{6}-2r_{\kk}^{6}\right)^2}S_{4}(r)\partial_{\mu}G_{E}(x^{\sigma}),
\end{align}
with eigenvalue equation \cite{Brower:2000rp}
\begin{equation} \label{GEmodeeq}
    \frac{\d}{\d r}\left(r^{7}-r\,r_{KK}^{6}\right)\frac{\d}{\d r}S_{4}(r)+\left(L^{4}M_{E}^{2}r^{3}+\frac{432r^{5}r_{\kk}^{12}}{\left(5r^{6}-2r_{KK}^{6}\right)^{2}}\right)S_{4}(r)=0.
\end{equation}
However, with $M_E=855$ MeV its mass is only a bit higher than that of the $\rho$ meson, whereas the predominantly dilatonic mode that is the
ground state of another tower of scalar modes with respect to 3+1 dimensions is only a little lighter than the traditional glueball candidates $f_0(1500)$ and $f_0(1710)$. This mode is degenerate with the tensor mode and involves
only metric fluctuations $\delta G_{11,11}$ and $\delta G_{\mu\nu}$, see
\eqref{eq:dilatonFluc}.

The exotic scalar glueball, denoted by $G_E$ in the following,
turns out \cite{Brunner:2015oqa} to have a relative width $\Gamma/M$ that is
much higher than that of the predominantly dilatonic scalar glueball ($G_D$),
but only the latter has a $\Gamma/M$ in the ballpark of $f_0(1500)$ and $f_0(1710)$. 

It was therefore proposed in \cite{Brunner:2015oqa} to
discard $G_E$ from the spectrum of glueballs of the WSS model as
a spurious mode that perhaps disappears in the inaccessible
limit $\MKK\to\infty$, where the supergravity approximation breaks down.
Already in \cite{Constable:1999gb} it was speculated that only one of the two
scalar glueball towers might correspond to the glueballs in QCD. Since it appears somewhat unnatural that an excited scalar glueball should have a smaller width than the ground-state scalar glueball, \cite{Brunner:2015oqa} preferred
the dilatonic scalar glueball as candidate for the actual ground state.

Indeed, the dilatonic scalar glueball turns out
to have a decay pattern that can match surprisingly well the
glueball candidate $f_0(1710)$,
in particular when including
additional couplings associated with the quark mass term
\cite{Brunner:2015yha,Brunner:2015oga}.
This may actually apply instead to $f_0(1770)$, which was proposed 
originally in \cite{BES:2004twe} as an additional $f_0$ resonance between 1700 and 1800 MeV and more recently in \cite{Sarantsev:2021ein} in radiative $J/\psi$ decays, where it appears dominantly as the most glue-rich resonance.\footnote{The next (2023) update of the PDG \cite{Workman:2022ynf} will
in fact include $f_0(1770)$ as a separate resonance (C.\ Amsler, private communication).}

The fact that the ratio $\Gamma_{f_0\to K\bar K}/\Gamma_{f_0\to \pi\pi}$ is significantly higher for $f_0(1710)$ \cite{Workman:2022ynf} (or for $f_0(1770)$ 
according to \cite{Sarantsev:2021ein}) than expected
from a flavor-symmetric glueball coupling can be attributed
to the fact that dilaton fluctuations couple naturally to quark
mass terms, similar to, but more pronounced than, in a model by Ellis and Lanik \cite{Ellis:1984jv}. There is therefore no need to invoke
the previous conjecture of chiral suppression of scalar glueball decay \cite{Carlson:1980kh,Sexton:1995kd,Chanowitz:2005du}, which
was questioned in \cite{Frere:2015xxa}.

In the following we shall mainly explore the consequences of this identification of the scalar glueball. In the radiative decay rates considered here, the explicit quark
masses will however not modify the (chiral) results for the couplings; they are only included in phase space factors.

We shall however need to make assumptions on how to extrapolate
to realistic glueball masses, which we describe in more detail
below. While the mass of $f_0(1710)$
is not too much above the original mass of $G_D$ in the WSS model,
larger extrapolations are required for the tensor and pseudoscalar glueballs when comparing to the various glueball candidates or lattice results.

\begin{table}
    \centering{}\bigskip{}
    \begin{tabular}{lcc}
    \toprule 
    $M_E$ & $\Gamma_{G_{E}}^{x=0}${[}MeV{]}&  $\Gamma_{G_{E}}^{x=1}${[}MeV{]} \tabularnewline\colrule
855 & 72\dots96 & 85\dots113 \tabularnewline
1506 & 286\dots383 & 430\dots570 \tabularnewline
1712 & 351\dots469 & 483\dots640 \tabularnewline
1865 & 398\dots530 & 521\dots691 \tabularnewline\botrule
    \end{tabular}
%
\qquad
    \begin{tabular}{lcc}
    \toprule 
    $M_D$ & $\Gamma_{G_{D}}^{x=0}${[}MeV{]}&  $\Gamma_{G_{D}}^{x=1}${[}MeV{]} \tabularnewline\colrule
1487 & 19\dots26 & 80\dots106 \tabularnewline
1506 & 19\dots27 & 80\dots106 \tabularnewline
1712 & 88\dots113 & 139\dots180 \tabularnewline
1865 & 151\dots197 & 198\dots259 \tabularnewline\botrule
    \end{tabular}   
    \caption{Total decay widths of the exotic and the dilatonic scalar glueball $G_E$ and $G_D$ with original masses of 855 MeV and 1487 MeV, respectively, and also with extrapolations to the masses of the glueball candidates $f_0(1510)$ or $f_0(1710)$ and the fragmented glueball of \cite{Sarantsev:2021ein,Klempt:2021nuf,Klempt:2021wpg}, for two choices of the extra coupling parameter $x$ associated with the quark mass term as defined in \cite{Brunner:2015oga}. The range of results corresponds again to $\lambda=16.63\dots12.55$. In addition to the two-body decays reviewed in Appendix \ref{app:hadrGBdecays}, the decays $G_D\to\rho\pi\pi\to 4\pi$ which interfere destructively with $G_D\to\rho\rho\to 4\pi$ have been taken into account here.
    }
    \label{tab:GDEtotalwidth}
\end{table}

As an alternative scenario, we shall also consider the
option of keeping the exotic scalar glueball mode $G_E$, whose relative
decay width $\Gamma/M$ is much too large to be identified with the traditional
glueball candidates $f_0(1510)$ or $f_0(1710)$
with total width 112(9) MeV and 128(18) MeV, respectively,
see Table \ref{tab:GDEtotalwidth}.
It would in fact fit better to the proposal in
\cite{Sarantsev:2021ein,Klempt:2021nuf,Klempt:2021wpg}
of a relatively broad fragmented glueball of mass
1865 MeV and a width of 370 MeV that does not show up as a
separate meson but only as admixture in the mesons
$f_0(1710)$, a novel $f_0(1770)$, $f_0(2020)$, and $f_0(2100)$.
Of course, this requires a drastic rise of the original mass of
$G_E$ by a factor of over 2, but also the mass of the
tensor mode $G_T$ would have to be raised
by a factor of 1.6 to match the expectation of $m_T\sim 2400$ MeV from lattice QCD; the mass of $G_D$, which would then be identified with
the first excited scalar glueball, would need to be raised somewhat
more, as lattice results point to a mass above the tensor glueball,
from around 2670 MeV\cite{Morningstar:1999rf}
to around 3760 MeV \cite{Gregory:2012hu}.

\subsection{Extrapolations to realistic glueball masses}
\label{sec:extrapol}

In the WSS model, the masses of glueballs are given by pure numbers times $\MKK$, which is also the case for the (axial) vector mesons.
However, when $\MKK$ has been fixed by the mass of the $\rho$ meson,
the glueball masses appear to be too small compared to lattice QCD results.

In order to predict decay rates for different glueball candidates we
manually change the masses of glueball modes in amplitudes and
phase space integrals, which could be viewed as assuming a different
scale $\MKK$ for the glueball sector.
The coupling constants involving glueballs
are all inversely proportional to $\MKK$ and we interpret
this appearance of $\MKK$ to be tied to the mass scale of glueballs,
which shows up also in their normalization factors $\mathcal{N}$, whereas
explicit appearances of $\MKK$ in the DBI action of the D branes
are considered as being fixed like the mass of the $\rho$ meson.
When upscaling glueball masses, we have therefore
correspondingly reduced the dimensionful 
glueball-meson/photon coupling constants.
[Without such a rescaling, the results for all glueball decay rates and the glueball contributions to $a_\upmu$ presented in Sect.~\ref{sec:rgd} extrapolated to some mass $M_G$ would be simply larger by a factor $(M_G/M_G^\mathrm{WSS})^2$.]

We consider this rescaling plausible in that the overlap
integrals of glueball and meson holographic profiles should
become smaller when glueball and meson modes are separated further in energy. It may well be, however, that this reduction is
only insufficiently accounted for by the overall change
of the mass scale in the glueball coupling constants;
thus our numerical results should be considered as somewhat rough estimates.

\section{Radiative Glueball Decays}
\label{sec:rgd}

In the following we shall concentrate on glueball interactions involving
vector mesons which through VMD also give rise to glueball-photon vertices.
Other hadronic interactions of glueballs are reviewed in Appendix \ref{app:hadrGBdecays}.

We shall consider the first three lightest glueball states, scalar, tensor, and
pseudoscalar 
in turn, choosing the dominantly dilatonic
scalar glueball over the exotic scalar glueball, since the former has been found to match remarkably well to the decay pattern of the glueball candidate $f_0(1710)$. The more unwieldy results for the exotic scalar glueball are
worked out in Appendix \ref{app:hadrGBdecays} and \ref{app:GE}.

\subsection{Dilatonic Scalar Glueball Decays}
\label{sec:GDdecays}


Inducing the fluctuation \eqref{eq:dilatonFluc} in the D8 brane action
\eqref{eq:9dAction} we obtain the interaction terms of the dilatonic scalar glueball
with two vector mesons as
\begin{align}
    \mathcal{L}_{G_{D} v^{m}v^{n}}= & \tr\int\d ^4x\,\left(d_{3}^{mn}\eta^{\rho\sigma}F_{\mu\rho}^{m}F_{\nu\sigma}^{n}+d_{2}^{mn}M_{\kk}^{2} v_{\mu}^{m} v_{\nu}^{n}\right)\left(\eta^{\mu\nu}-\frac{\partial^{\mu}\partial^{\nu}}{\Box}\right)G_{D},\label{eq:dilScalarVectorInteraction}
\end{align}
where the coupling constants are given by 
\begin{equation}
\begin{split}\label{d2d3}
        &d_{2}^{mn}= \kappa\int\d z\,K\psi_{2n-1}^{\prime}\psi_{2m-1}^{\prime}H_{D}=\{4.3714,\dots\}\frac{1}{\lambda^{\frac{1}{2}}N_{c}M_{\kk}},\\
        &d_{3}^{mn}=\kappa\int\d z\,K^{-1/3}\psi_{2n-1}\psi_{2m-1}H_{D}=\{18.873,\dots\}\frac{1}{\lambda^{\frac{1}{2}}N_{c}M_{\kk}}.
\end{split}
\end{equation}    

Restricting to the ground-state vector mesons ($m=n=1$), the amplitudes for
the decay of the dilatonic scalar glueball into vector mesons with transverse and longitudinal polarizations read
\begin{align}
    \left|\mathcal{M}_{T}^{\left(G_{D}\rightarrow v^{1}v^{1}\right)}\right| & =\left[d_{3}^{11}\left(2m_{v^{1}}^{2}-\frac{3M_{D}^{2}}{4}\right)-d_{2}^{11}M_{\kk}^{2}\right],\nonumber \\
    \left|\mathcal{M}_{L}^{\left(G_{D}\rightarrow v^{1}v^{1}\right)}\right| & =\left[\frac{d_{2}^{11}M_{D}^{2}M_{\kk}^{2}}{4m_{v^{1}}^{2}}+d_{3}^{11}m_{v^{1}}^{2}\right],
\end{align}
in terms of which the partial decay width is given by
\begin{align}\label{GammaGDvv}
    \Gamma_{G_{D}\rightarrow v^{1,a}v^{1,b}} & =\frac{1}{S}\left(2\left|\mathcal{M}_{T}^{\left(G_{D}\rightarrow v^{1}v^{1}\right)}\right|^{2}+\left|\mathcal{M}_{L}^{\left(G_{D}\rightarrow v^{1}v^{1}\right)}\right|^{2}\right)\frac{\left|\mathbf{p}_{v^{1}}\right|}{8\pi M_{D}^2}, 
\end{align}
where $S$ equals 2 for identical particles ($a=b$) and 1 otherwise.

In the narrow-resonance approximation, this vanishes for the WSS model mass $M_D=1487$ MeV, which is below the threshold of two $\rho$ mesons.
However,
when $M_D$ is manually adjusted to the mass of $f_0(1710)$,
which we assume as 1712 MeV (the average of the $T$-matrix pole results of \cite{WA102:1999fqy} and \cite{WA102:2000lao}), the decay $G_D\to\rho\rho$
becomes the largest channel, exceeding even the dominant pseudoscalar
channel $G_D\to KK$ (see Appendix \ref{app:hadrGBdecays}, Table \ref{tab:GDhadronicdecays}).

As discussed in \BPR, the holographic prediction for the 
total rate $G_D\to 4\pi$ is somewhat reduced by a destructive interference
from $G_D\to\rho\pi\pi$, rendering the partial width of $G_D\to 4\pi$ similar to and slightly less than $G_D\to KK$ \cite{Brunner:2015yha}.
Remarkably, data from radiative $J/\psi$ decays \cite{BES:1999dmf} for $f_0(1740)$ (or $f_0(1770)$ in \cite{Sarantsev:2021ein})
seem to be fairly consistent with this result.

\subsubsection{Dilatonic scalar glueball \texorpdfstring{$1\gamma$-decays}{1 photon}}
\label{subsec:GD1gamma}

From the interaction terms \eqref{eq:dilScalarVectorInteraction} we
can also derive the interactions including photons by using VMD.
Replacing one vector meson by a photon we find
\begin{align}
    \mathcal{L}_{G_{D}\V v^{m}}= & 2d_{3}^{m\V}\eta^{\rho\sigma}\tr\left(F_{\mu\rho}^{m}F_{\nu\sigma}^{\V}\right)\left(\eta^{\mu\nu}-\frac{\partial^{\mu}\partial^{\nu}}{\Box}\right)G_{D},
\end{align}
with 
\begin{align}\label{d3mV}
    d_{3}^{m\V}\equiv & \kappa\int\d z\,K^{-1/3}\psi_{2m-1}H_{D}\nonumber\\
    &=\left\{ 0.46895,\dots\right\} \frac{1}{M_{\kk}\sqrt{N_{c}}}.
\end{align}
The other coupling $d_{2}^{m\V}$ vanishes for an on-shell photon,
since at zero virtuality its radial mode is constant and drops 
out in the replacement $\psi'\to\mathcal{J}'=0$. 

In radiative decays,
only the transverse amplitude remains, which reads
\begin{align}
    \left|\mathcal{M}_{T}^{\left(G_{D}\rightarrow\V v^{m}\right)}\right| & =\frac{d_{3}^{m\mathcal{V}}\left(m^{4}_{v}-4m^{2}_{v}M_{D}^{2}+3M_{D}^{4}\right)}{2M_{D}^{2}} \tr\left( e Q T_{v}\right),
\end{align}
yielding
\begin{align}\label{GammaGDvV}
    \Gamma_{G_{D}\rightarrow v^{m}\gamma} & =2\left|\mathcal{M}_{T}^{\left(G_{D}\rightarrow\V v^{m}\right)}\right|^{2}\frac{\left|\mathbf{p}_{\V}\right|}{8\pi M_{D}^2}.
\end{align}
The results
are displayed in Table \ref{tab:radGDDecayideal}
for two mass parameters corresponding to $f_0(1500)$ and $f_0(1710)$, where ideal mixing was assumed for the $\omega$ and $\phi$ mesons. The latter implies that
$\rho\gamma$ and $\omega\gamma$ decay rates are very close to the ratio $9:1$. 
The ratio of decay rates $\phi\gamma$
and $\omega\gamma$, which would be 2:1 with equal masses, is, however, significantly reduced by the larger $\phi$ mass.\footnote{A more realistic value for the $\phi$-$\omega$ mixing angle of 3.32$^\circ$ above ideal mixing \cite{Ambrosino:2009sc} increases the partial width for $\omega\gamma$ by about 17\% and decreases the one for $\phi\gamma$ by about 8.5\%. This also holds true for all the other glueball decay widths below.\label{fn-omegaphi}}

\subsubsection{Dilatonic scalar glueball \texorpdfstring{$2\gamma$-decays}{2 photon}}

Replacing the second vector meson by a photon by means of VMD, we obtain the $2\gamma$ interactions
\begin{align}
    \mathcal{L}_{G_{D}\V\V}= & d_{3}^{\V\V}\eta^{\rho\sigma}\tr\left(F_{\mu\rho}^{\V}F_{\nu\sigma}^{\V}\right)\left(\eta^{\mu\nu}-\frac{\partial^{\mu}\partial^{\nu}}{\Box}\right)G_{D},
\end{align}
with 
\begin{align}\label{d3VV}
    d_{3}^{\mathcal{V}\mathcal{V}}\equiv & \kappa\int\d z\,K^{-1/3}H_{D}=0.0130195\lambda^{1/2}M_{\kk}^{-1}
\end{align}
which gives
\begin{align}
        \left|\mathcal{M}_T^{\left(G_{D}\rightarrow\V\V\right)}\right| & =\frac{3}{2}d_{3}^{\V\V} M_{D}^{2}\tr( e^2 Q^2 )
\end{align}

The resulting width
\begin{align}\label{GammaGDVV}
    \Gamma_{G_{D}\rightarrow\gamma\gamma} & =\frac{1}{8\pi}\left|\mathcal{M}_T^{\left(G_{D}\rightarrow\V\V\right)}\right|^{2}\frac{\left|\mathbf{p}_{\V}\right|}{M_{D}^2}
\end{align}
is again displayed in Table \ref{tab:radGDDecayideal} for
the two mass parameters corresponding to $f_0(1500)$ and $f_0(1710)$,
which in both cases is above 1 keV.

This is larger than the old prediction by Kada et al.\ \cite{Kada:1988rs},
but an order of magnitude smaller than the VMD based result
of Cotanch and Williams
\cite{Cotanch:2005ja}, who obtained 15.1 keV for a scalar glueball with
mass 1700 MeV after correcting their previous result of 2.6 keV in \cite{Cotanch:2004py} (note that the corresponding preprint has erroneously 2.6 eV instead). Also all other radiative decay rates obtained in \cite{Kada:1988rs} are about an order of magnitude larger than ours (not uniformly so, however, but varying between a factor of 7 to 26, thereby deviating from the ratios discussed at the end of Sect.~\ref{subsec:GD1gamma}).

\begin{table}
    \centering{}\bigskip{}
    \begin{tabular}{lc}
    \toprule 
    &  $\Gamma_{G_{D}}${[}keV{]}\tabularnewline\colrule
    $f_{0}\left(1500\right)\rightarrow\rho\gamma$ & 184\tabularnewline
    $f_{0}\left(1500\right)\rightarrow\omega\gamma$ & 19.9\tabularnewline
    $f_{0}\left(1500\right)\rightarrow\phi\gamma$  & 14.1\tabularnewline
$f_{0}\left(1500\right)\rightarrow\gamma\gamma$ & 1.74\dots1.32\tabularnewline\colrule
    $f_{0}\left(1710\right)\rightarrow\rho\rho$ & (53.5\dots 71.0)$\cdot10^{3}$\tabularnewline
    $f_{0}\left(1710\right)\rightarrow\omega\omega$ & (16.6\dots22.0)$\cdot10^{3}$\tabularnewline
    $f_{0}\left(1710\right)\rightarrow\rho\gamma$  & 276\tabularnewline
    $f_{0}\left(1710\right)\rightarrow\omega\gamma$ & 30.1\tabularnewline
    $f_{0}\left(1710\right)\rightarrow\phi\gamma$  & 29.4\tabularnewline
    $f_{0}\left(1710\right)\rightarrow\gamma\gamma$ & 1.98\dots1.50\tabularnewline\botrule
    \end{tabular}\caption{Radiative scalar glueball decay with $G_D$ identified alternatively with $f_0(1500)$ and $f_0(1710)$ with masses 1506 MeV and 1712 MeV, respectively, for $\lambda=16.63\dots12.55$.}
    \label{tab:radGDDecayideal}
\end{table}

On the other hand, the two-vector meson decay rates obtained in
\cite{Cotanch:2004py} (44.4 MeV for $\rho\rho$ and 34.6 MeV for $\omega\omega$) are not very far from our results.
In fact, our holographic prediction for $f_0(1710)\to\omega\omega$
with $f_0(1710)$ as a (predominantly dilaton) glueball
appears to be in the right ballpark considering the measured branching ratios
of radiative $J/\psi$ decays in $\gamma f_0(1710)\to \gamma K\bar K$ and $\gamma f_0(1710)\to \gamma \omega\omega$ \cite{Workman:2022ynf}
(which according to \cite{Sarantsev:2021ein} may be instead $f_0(1770)$). 
The PDG \cite{Workman:2022ynf} quotes two results for $\mathcal{B}(K\bar K)$: a BNL measurement \cite{Longacre:1986fh} from 1986 with
$\mathcal{B}(K\bar K)=0.38^{+0.09}_{-0.19}$ and a phenomenological analysis \cite{Albaladejo:2008qa} concluding $0.36(12)$, which both are consistent with the WSS result obtained in \cite{Brunner:2015yha} as approximately 0.35.
Using $\mathcal{B}(K\bar K)=0.36(12)$ and the total decay width of $f_0(1710)$ \cite{Workman:2022ynf} of 123(18) MeV lead to a partial decay width 
for $f_0(1710)\to \omega\omega$ of about 15(8) MeV, for which the holographic prediction
from $G_D$ amounts to $16.6\dots22.0$ MeV.

No experimental results for single-photon decays of $f_0(1710)$ appear to be available, but
in \cite{Belle:2013eck} the BELLE collaboration reports a measurement of
$f_{0}(1710)\rightarrow\gamma\gamma$ with the result
$\Gamma_{\gamma\gamma}\mathcal{B}(K\bar K)=12_{-2-8}^{+3+227}\;\mathrm{eV}$,
with the stated conclusion that the $f_0(1710)$ meson was
unlikely to be a glueball because of a width larger than that expected (``much less than 1 eV'') for a pure glueball state. However the holographic prediction for $\Gamma_{\gamma\gamma}\mathcal{B}(K\bar K)\approx 690\dots520\;\mathrm{eV}$ is 3-2 $\sigma$ above the upper limit of the BELLE result.\footnote{Older upper limits for $\Gamma_{\gamma\gamma}\mathcal{B}(K\bar K)$ are
480 eV from ARGUS \cite{ARGUS:1989ird}, 200 eV from CELLO \cite{CELLO:1988xbx}, and 560 eV from TASSO \cite{TASSO:1985tme}. (The latter two are quoted by the PDG \cite{Workman:2022ynf} with lower values, 110 eV and 280 eV, respectively, corresponding however to the assumption of helicity 2 which leads to smaller upper limits.)}
Ironically, the BELLE result for the two-photon rate appears to be rather too small for a pure (predominantly dilaton) glueball interpretation of  $f_{0}(1710)$ within the WSS model.\footnote{Assuming a tensor glueball $f_2(1720)$, \cite{Kada:1988rs} predicted $\Gamma_{\gamma\gamma}\mathcal{B}(K\bar K)\approx 95$ eV.}
The central value of the BELLE result for $\Gamma_{\gamma\gamma}$ of only a few tens of eV would thus seem to indicate that VMD does not apply for radiative decays of $f_0(1710)$.

In Appendix \ref{app:GE} we also evaluate radiative decays of the exotic glueball of the WSS model. The two-photon decay width of $G_E$ is considerably smaller than that of $G_D$, 87\dots65 eV, when the mass of $G_E$ is extrapolated to that of $f_0(1710)$. However, the decay pattern of $G_E$ does not fit to either $f_0(1500)$ or $f_0(1710)$ when extrapolating to their masses.

\subsection{Tensor Glueball Decays}
\label{sec:GTdecays}

The holographic mode functions associated with tensor glueballs are reviewed in Appendix \ref{app:tensor} together with the results of hadronic two-body decays.

Radiative decays of tensor glueballs can be derived from
the interaction terms with two vector mesons, which are given by
\begin{align}
    \mathcal{L}_{G_{T} v^{m} v^{n}}= & \tr\left[t_{2}M_{\kk}^{2} v_{\mu}^{m} v_{\nu}^{n}G_{T}^{\mu\nu}+t_{3}F_{\mu\rho}^{m}F_{\nu}^{n\rho}G_{T}^{\mu\nu}\right],\label{eq:tensormesondecay}
\end{align}
with
\begin{align}
    t_{2}^{mn}= & \int\d zK\psi_{2m-1}^{\prime}\psi_{2n-1}^{\prime}T=2\sqrt{3}d_{2}^{mn}.
\end{align}
\begin{align}
    t_{3}^{mn}= & \int\d zK^{-1/3}\psi_{2m-1}\psi_{2n-1}T=2\sqrt{3}d_{3}^{mn}
\end{align}
and $d_{2,3}^{mn}$ as given in \eqref{d2d3}.
(Note that due to a different normalization of the tensor field, the tensor coupling constants differ from those in \BPR\ by a factor $\sqrt2$; all other glueball coupling constants are defined as in \BPR.)

The decay rate of a tensor glueball into two vector mesons reads
\begin{align}
\Gamma_{G_T\to vv}=\frac{1}{S}
\biggl\{
&\frac{t_2^2}{120}\frac{M_\kk^4}{m_v^4}(M_G^4+12m_v^2 M_G^2+56m_v^4)\nonumber\\
&+\frac{2}{3}t_2 t_3 M_\kk^2 (M_G^2-m_v^2)\nonumber\\
&+\frac{t_3^2}{10}(M_G^4-3m_v^2M_G^2+6m_v^4)
\biggr\}
\frac{\left|\mathbf{p}_{v}\right|}{8\pi M_{G}^2}, 
\end{align}
where $S$ is again the symmetry factor for identical particles.

\subsubsection{Tensor glueball \texorpdfstring{$1\gamma$-decays}{1 photon}}

Through VMD \eqref{eq:tensormesondecay} leads to a coupling of the tensor glueball with one photon and one vector meson with interaction Lagrangian
\begin{align}
    \mathcal{L}_{G_{T} v^{n}\V}= & 2t_{3}^{\V n}G_{T}^{\mu\nu}\eta^{\rho\sigma}\tr\left(F_{\mu\rho}^{\V}F_{\nu\sigma}^{n}\right),
\end{align}
with 
\begin{align}
    t_{3}^{\V n}= & \int\d zK^{-1/3}\psi_{2n-1}T=2\sqrt{3}d_{3}^{\mathcal{V}n}
\end{align}
and $d_{3}^{\mathcal{V}n}$ as given in \eqref{d3mV}.

This yields
\begin{equation}
    \Gamma_{G_T\to v\gamma}=
    \frac{1}{15M_G^4}\left(\tr\left(eQT_v\right)\right)^2\left(M_G^2-m_v^2\right)^2\left(6M_G^4+3M_G^2m_v^2+2m_v^4\right)
    \frac{|\mathbf{p}_v|}{8\pi M_G^2}. 
\end{equation}

\subsubsection{Tensor glueball \texorpdfstring{$2\gamma$-decays}{2 photon}}

Similarly \eqref{eq:tensormesondecay} leads to
\begin{align}
    \mathcal{L}_{G_{T}\V\V}= & t_{3}^{\V\V}G_{T}^{\mu\nu}\eta^{\rho\sigma}\tr\left(F_{\mu\rho}^{\V}F_{\nu\sigma}^{\V}\right),
\end{align}
with 
\begin{align}
    t_{3}^{\V\V}= & \int\d zK^{-1/3}T=2\sqrt{3}d_{3}^{\V\V}.
\end{align}
and $d_{3}^{\V\V}$ as given in \eqref{d3VV}.

The resulting two-photon decay width of the tensor glueball is given by
\begin{equation}
    \Gamma_{G_T\to\gamma\gamma}=\frac{1}{5}\left[t_{3}^{\V\V}M_{G}^{2}  \tr\left( e^2 Q^2\right)\right]^2
    \frac{|\mathbf{p}_\gamma|}{8\pi M_G^2}.
\end{equation}

The resulting partial widths are listed in Table \ref{radTensorDecay} for three values of the mass of the tensor glueball, the unrealistically small WSS model mass value 1487 MeV as well as two higher values motivated
by pomeron physics \cite{Donnachie:2002en}\footnote{A candidate for a tensor glueball around 2000 MeV is the broad resonance $f_2(1950)$, which has recently also been argued for in \cite{Vereijken:2023jor} on the basis of a chiral hadronic model. The latter turns out to yield a dominance of the decay modes into two vector mesons, in qualitative agreement with the WSS model, which in fact predicts a very broad tensor glueball (see Appendix \ref{app:tensor}).}
and QCD lattice studies \cite{Morningstar:1999rf}, respectively,
assuming ideal mixing of $\omega$ and $\phi$ mesons. 
With increasing mass of the glueball, the partial decay widths for
$\rho\gamma$, $\omega\gamma$, and $\phi\gamma$ gradually approach the
ratios $9:1:2$ for degenerate vector meson masses; again, a more realistic value of $\theta_V$ changes the $\omega\gamma$ and $\phi\gamma$ results only slightly {(cf.\ footnote \ref{fn-omegaphi})}.

The radiative decay widths obtained for the tensor glueball
turn out to be comparable with those for the dilatonic scalar glueball for equal glueball mass, rising approximately linear
with glueball mass (due to the rescaling described in Sect.\ \ref{sec:extrapol}).

Our prediction of the two-photon width of $\sim$ 2-3 keV
is significantly larger than the old prediction of Kada et al.\ \cite{Kada:1988rs} who have
values in the range of hundreds of eV, 
and also higher than the more recent prediction in
\cite{Godizov:2016vuw}, where $\Gamma_{f_2(1950)\to\gamma\gamma}=960(50)$ eV was obtained.
Cotanch and Williams \cite{Cotanch:2005ja}, on the other hand, have
also results above 1 keV, $\Gamma_{G_T(2010)\to\gamma\gamma}=1.72$ keV
and $\Gamma_{G_T(2300)\to\gamma\gamma}=1.96$ keV, by using VMD.
Also their results for single-photon decays are comparable with ours,
even though their results for decays into two vector mesons are
significantly smaller than ours. A particular point of disagreement
is their result for a relatively large $\omega\phi$ decay mode,
which in the WSS model is absent. As noted in \cite{Giacosa:2005bw}, this is possible only by allowing for a rather strong deviation from the large-$N_c$ limit.

\begin{table}
    \centering{}\bigskip{}
    \begin{tabular}{lccc}
    \toprule
    & $\Gamma_{G_{T}^\mathrm{WSS}}${[}keV{]} & $\Gamma_{G_{T}(2000)}${[}keV{]} & $\Gamma_{G_{T}(2400)}${[}keV{]} \tabularnewline\colrule    
    $G_{T}\rightarrow
    \rho\rho$   & - & (270\dots358)$\cdot 10^{3}$ & (382\dots507)$\cdot 10^{3}$ \tabularnewline
    $G_{T}\rightarrow
    \omega\omega$  & - & (88.2\dots117)$\cdot 10^{3}$ & (127\dots169)$\cdot10^{3}$ \tabularnewline 
    $G_{T}\rightarrow 
    K^{*}K^{*}$  & - & (240\dots318)$\cdot 10^{3}$ & (417\dots552)$\cdot 10^{3}$ \tabularnewline
    $G_{T}\rightarrow
    \phi\phi$  & - & - & (76.7\dots102)$\cdot 10^{3}$ \tabularnewline\colrule  
    $G_{T}\rightarrow
    \rho\gamma$    & 260 & 522 & 716\tabularnewline
    $G_{T}\rightarrow
    \omega\gamma$    & 28.3  & 57.5 & 79.1\tabularnewline
    $G_{T}\rightarrow
    \phi\gamma$    & 24.7 & 81.1 & 127\tabularnewline\botrule  
    $G_{T}\rightarrow
    \gamma\gamma$    & 1.84\dots1.39 & 2.47\dots 1.86 & 2.97\dots2.24\tabularnewline\botrule 
    \end{tabular}\caption{Radiative tensor glueball decays and decays into two vector mesons for $\lambda=16.63\dots12.55$. Besides the pristine results for the WSS model mass of 1487 MeV, their extrapolations to glueball masses of 2000 and 2400 MeV are given.}
    \label{radTensorDecay} 
\end{table}

\subsection{Pseudoscalar Glueball Decays}

In the WSS model, the pseudoscalar glueball is represented by a Ramond-Ramond 1-form field $C_1$, which has a kinetic mixing with the singlet $\eta_0$
given by \cite{Leutgeb:2019lqu}
\be\label{GPSmixing}
\eta_0\to\eta_0+\zeta_2\, G_{PS}=
\eta_0+0.01118 \sqrt{N_f/N_c}\, \lambda\, G_{PS},
\ee
with $G_{PS}$ remaining unchanged to leading order in $\sqrt{N_f/N_c}$
(formally treated as a small quantity because of the probe brane approximation).
In contrast to the conventional mixing scenarios of Ref.~\cite{Rosenzweig:1981cu,Mathieu:2009sg} mass mixing is absent here, while the mass of the pseudoscalar glueball is raised by a factor $(1+\zeta_2^2)$ from 1789 MeV to (1819.7\dots 1806.5) MeV for $\lambda=16.63\dots12.55$. Lattice QCD (in the quenched approximation), however, typically finds
values around 2600 MeV, so we also consider the latter in our extrapolations.\footnote{Note that historically the pseudoscalar glueball was expected to be the lightest glueball, with $\eta(1405)$ a prominent candidate after $\iota(1440)$ \cite{Edwards:1982nc} was split into $\eta(1405)$ and $\eta(1475)$. This is still occasionally considered a possibility, see for example \cite{Masoni:2006rz} and \cite{Cheng:2008ss}.
}

Through \eqref{GPSmixing} the pseudoscalar glueball acquires the same interactions as $\eta_0$, and
the same form of transition form factors, only with correspondingly
modified coupling constants.
Thus the formulae given in \ref{sec:raddecpscvm} for the decays of pseudoscalars in vector mesons or photons remain essentially unchanged, but the higher mass of the pseudoscalar glueball permits also decays into pairs of vector mesons.

The resulting interaction Lagrangian reads
\be
    \mathcal{L}_{G_{PS}vv/v\mathcal{V}/\mathcal{V}\mathcal{V}}= G_{PS}\epsilon^{\mu\nu\rho\sigma}\tr
    \left[k_{1}^{vv}\partial_{\mu} v_{\nu}\partial_{\rho}v_{\sigma}
    +2k_{1}^{v\mathcal{V}}\partial_{\mu} v_{\nu}\partial_{\rho}\mathcal{V}_{\sigma}
    +k_{1}^{\mathcal{V}\mathcal{V}}\partial_{\mu} \mathcal{V}_{\nu}\partial_{\rho}\mathcal{V}_{\sigma}
    \right]
\ee
with\footnote{The couplings differ by a factor of 2 from \cite{Leutgeb:2019lqu} since we use SU(3) generators 
$T^a=\lambda^a/2$.}
\begin{align}
    k_{1}^{v^{1}v^{1}}
    &=  19.6184N_{c}^{-1}\lambda^{-1/2}M_{\kk}^{-1},\\
    k_{1}^{v^{1}\V}
    &=  0.493557N_{c}^{-1/2}M_{\kk}^{-1},\\
    k_{1}^{\V\V}
    &=  0.0145232\lambda^{1/2}M_{\kk}^{-1}.
\end{align}
The various resulting partial widths are listed in Table \ref{radPSGDecay}.

In the WSS model, all other hadronic decay channels of the pseudoscalar glueball, such as those considered in \cite{Eshraim:2012jv,Eshraim:2016mds}, turn out to be very weak compared to two-vector-meson decays \cite{Leutgeb:2019lqu}.
The relative strength of the latter entails correspondingly
important radiative decay modes, and a two-photon partial width
in the keV range. Note, however, that these results have been
obtained from the first term in a formal expansion in $\sqrt{N_f/N_c}$, which is not a small parameter in real QCD. It
might nevertheless be meaningful, since
the parameter $\zeta_2$ in \eqref{GPSmixing} is reasonably small,
0.19\dots0.14 for $\lambda=16.63\dots12.55$.

\begin{table}
    \centering{}\bigskip{}  
    \begin{tabular}{lcc}
    \toprule
    & $\Gamma_{G_{PS}^\mathrm{WSS}}${[}keV{]} & $\Gamma_{G_{PS}(2600)}${[}keV{]}\tabularnewline\colrule  
    $G_{PS}\rightarrow\rho\rho$ & (36.8\dots45.0)$\cdot10^{3}$ & (190\dots248)$\cdot10^{3}$\tabularnewline
    $G_{PS}\rightarrow\omega\omega$ & (11.3\dots13.8)$\cdot10^{3}$ & (62.2\dots81.3)$\cdot10^{3}$\tabularnewline
    $G_{PS}\rightarrow\phi\phi$ & - & (29.2\dots38.2)$\cdot10^{3}$\tabularnewline
    $G_{PS}\rightarrow K^{*}K^{*}$ & (2.69\dots1.81)$\cdot10^{3}$ & (188\dots246)$\cdot10^{3}$\tabularnewline\colrule
    $G_{PS}\rightarrow\rho\gamma$ & 272\dots263 & 536\dots528\tabularnewline
    $G_{PS}\rightarrow\omega\gamma$ & 29.8\dots28.9 & 59.2\dots58.3\tabularnewline
    $G_{PS}\rightarrow\phi\gamma$ & 35.6\dots34.1 & 95.4\dots94.0\tabularnewline
    \hline
    $G_{PS}\rightarrow\gamma\gamma$ & 1.75\dots 1.30 & 2.49\dots1.86\tabularnewline\botrule
    \end{tabular}\caption{Radiative pseudoscalar glueball decay and decays into two vector mesons $\lambda=16.63\dots12.55$. Besides the WSS model result for the pseudoscalar mass, $M_{G}=1813\pm7\text{MeV}$, an extrapolation to 2600 MeV (motivated by lattice results) is considered.}
    \label{radPSGDecay}
\end{table}

\section{Glueball contributions to hadronic light-by-light scattering and the muon \texorpdfstring{$g-2$}{g-2}}

In order to calculate the contribution of the glueball exchange diagram in the light-by-light scattering amplitude, which enters the muon-photon vertex at two loop order, the above results for the vertices of a glueball with two on-shell photons need to be generalized to nonzero photon virtualities.

In the case of the dilatonic scalar glueball $G_D$, this involves two
interaction terms that are obtained by replacing $v_\mu$ in \eqref{eq:dilScalarVectorInteraction} by
$eQA_\mu^{e.m.}$ and the holographic profile functions $\psi(z)$ in \eqref{d2d3} by
the bulk-to-boundary propagator $\mathcal{J}(Q,z)$ defined in \eqref{JQz}, yielding two form factors,
\begin{align}
    d_{2}^{\mathcal{V}\mathcal{V}}(Q_1^2,Q_2^2) &\equiv \kappa \int dz\,K \partial_z\mathcal{J}(Q_1,z) \partial_z\mathcal{J}(Q_2,z)  H_{D}(z),\nonumber\\
    d_{3}^{\mathcal{V}\mathcal{V}}(Q_1^2,Q_2^2) &\equiv \kappa \int dz\,K^{-1/3} \mathcal{J}(Q_1,z) \mathcal{J}(Q_2,z)  H_{D}(z),
\end{align}
in place of the coupling constants $d_2$ and $d_3$.

The exotic scalar glueball $G_E$ has more complicated interactions with two vector fields, written out in \eqref{LGEvv}, with five coupling constants \eqref{c234}.
The latter are generalized in a completely analogous manner to
form factors $c_i^{\mathcal{V}\mathcal{V}}(Q_1^2,Q_2^2)$ with $i=2,3,4$, and
$\breve{c}_j^{\mathcal{V}\mathcal{V}}(Q_1^2,Q_2^2)$ with $j=2,3$.

Following the notation of \cite{Danilkin:2021icn}, the result for the matrix element of a scalar glueball with two electromagnetic currents $j_\text{em}^\mu(x)$ can be written in terms of two transition form factors $\mathcal{F}_{1,2}^S$ defined by
\begin{eqnarray}
    \mathcal{M}^{\mu \nu}(p\to q_1,q_2)&=& i \int d^4 x e^{i q_1\cdot x} \langle 0 \left| j_\text{em}^\mu(x) j_\text{em}^\nu(0) \right| G_S(p) \rangle \nonumber \\
    &=& \frac{\mathcal{F}_1^S(q_1^2,q_2^2)}{M_S} T_1^{\mu \nu} +\frac{\mathcal{F}_2^S(q_1^2,q_2^2)}{M_S^3} T_2^{\mu \nu}
\end{eqnarray}
with 
\begin{eqnarray}
    T_1^{\mu\nu} &=& q_1\cdot q_2 g^{\mu\nu} - q_2^\mu q_1^\nu ,              \nonumber \\
    T_2^{\mu\nu} &=& q_1^2 q_2^2 g^{\mu\nu} + q_1\cdot q_2 q_1^\mu q_2^\nu  - q_1^2 q_2^\mu q_2^\nu - q_2^2 q_1^\mu q_1^\nu.
\end{eqnarray}

For the dilatonic scalar glueball we obtain
\begin{align}
    \mathcal{F}_1^D&=-2\frac{d_3^{\mathcal{VV}}(Q_1^2,Q_2^2)\tr  Q^2}{M_D}\left[(q_1^2+q_2^2)+(q_1\cdot q_2)+2M_D^2)\right]-\frac{d_2^{\mathcal{VV}}(Q_1^2,Q_2^2)\MKK^2 \tr Q^2}{M_D},\\
    \mathcal{F}_2^D&=-2d_3^{\mathcal{VV}}(Q_1^2,Q_2^2)\tr Q^2 M_D+\frac{d_2^{\mathcal{VV}}(Q_1^2,Q_2^2) \MKK^2 \tr Q^2 M_D}{q_1^2q_2^2}\left[(q_1\cdot q_2)+M_D^2)\right],
\end{align}
and for the exotic scalar glueball
\begin{align}
    \mathcal{F}_1^E&= -2\frac{\tr Q^2}{M_E}\left[ c_{3}^{\V\V}(Q_1^2,Q_2^2)((q_1^2+q_2^2)+(q_1\cdot  q_2)+M_E^2)- \breve{c}_{3}^{\V\V}(Q_1^2,Q_2^2) M_E^2\right. \nonumber\\ &\left. +c_2^{\V\V}(Q_1^2,Q_2^2) \MKK^2-\frac{3}{2}\left(c_4^{\V\V}(Q_1^2,Q_2^2)+c_4^{\V\V}(Q_2^2,Q_1^2)\right)\right],\\
    \mathcal{F}_2^E&= -2\tr Q^2 M_E\left[ c_{3}^{\V\V}(Q_1^2,Q_2^2)-c_2^{\V\V}(Q_1^2,Q_2^2)\MKK^2\frac{(q_1\cdot q_2)}{q_1^2q_2^2}\right. \nonumber\\ & \left. + \breve{c}_2^{\V\V}(Q_1^2,Q_2^2)\frac{M_E^2\MKK^2}{q_1^2q_2^2}-\frac{3}{2}\MKK^2\frac{c_4^{\V\V}(Q_1^2,Q_2^2)q_1^2+c_4^{\V\V}(Q_2^2,Q_1^2)q_2^2}{q_1^2q_2^2}\right],
\end{align}
where $\q1\cdot \q2=-\frac{1}{2}(\q1^2+\q2^2+M_{D/E})$
and $\tr Q^2=2/3$ for $N_f=3$.

We have used these results to estimate the glueball contribution to the muon 
anomalous magnetic moment $a_\upmu=(g-2)_\upmu/2$ in a narrow-width approximation by inserting the above expressions in the two-loop
expression for the muon-photon vertex.

In the scenario where the exotic scalar glueball is discarded from the spectrum
and $G_D$ is identified with the ground-state scalar glueball,
we obtain for $M_D=1506$ MeV and $M_D=1712$ MeV corresponding to the
glueball candidates $f_0(1500)$ and $f_0(1710)$
\begin{align}
a_\upmu^{G_D(1506)}&=-1.62\times 10^{-12},\nonumber\\
a_\upmu^{G_D(1712)}&=-1.01\times 10^{-12}.
\end{align}
While the former result is approximately identical to the unmodified
WSS result, since $M_D^\mathrm{WSS}=1487$ MeV, the latter depends on the
specific extrapolations laid out in Sect.~\ref{sec:extrapol}. Had we
only raised the mass, it would have been somewhat larger, $-1.35\times 10^{-12}$, but in this case the rather good agreement of the hadronic decay pattern
obtained for $G_D(1712)$ with the experimental results for the glueball candidate $f_0(1710)$ (or $f_0(1770)$ according to \cite{Sarantsev:2021ein}) would
have deteriorated.

If the exotic scalar glueball is not discarded from the spectrum but
identified with the ground-state scalar glueball, its mass needs to be
raised substantially to match the predictions from lattice QCD.
Its decay pattern and in particular its large width then does not fit
to either $f_0(1500)$ and $f_0(1710)$; it might instead be identified
with the broad ``fragmented'' glueball $G(1865)$ proposed in \cite{Sarantsev:2021ein,Klempt:2021nuf,Klempt:2021wpg}.
Raising the mass of $G_E$ artificially to this glueball, we obtain
for its $a_\upmu$ contribution
\be\label{amuGE1865}
a_\upmu^{G_E(1865)}=-0.10\times 10^{-12},
\ee
which is an order of magnitude smaller in accordance with the
much smaller two-photon rate of $G_E$. Since in this case the narrow-width
approximation is rather questionable, we have also considered
the space-like Breit-Wigner function proposed in \cite{Knecht:2018sci}.
However, this changes the result \eqref{amuGE1865} only by about 2\%.

In \cite{Knecht:2018sci} the authors consider scalar resonances including $f_0(1500)$, which is assumed to have a sizeable photon coupling while being a glueball-like state, with a coupling constant similar to the one obtained for $f_0(980)$, leading to $\Gamma^{f_0(1500)\to\gamma\gamma}\approx 0.79\,\text{keV}$. The assumed transition form factors therein yield $a_\upmu=-(1.3\dots2)\times 10^{-12}$. This is comparable to our results, even though the two-photon rate obtained with $G_D$ is about twice as large.

In the WSS model, tensor glueballs have two-photon decay rates comparable to $G_D$ with similar values of $\Gamma_{\gamma\gamma}/M_G$. We have not evaluated their contribution to $a_\upmu$, but we expect that they will be 
smaller than those of $G_D$ by some power of the ratio
$M_{G_T}/M_{G_D}$.

We have however evaluated the contribution of
pseudoscalar glueballs, which contribute with a positive sign.
With the WSS model mass of
$1789$ MeV we find $a_\upmu^{G_{PS}^\mathrm{WSS}}=0.39\times10^{-12}$,
and when extrapolated to a value typically found in quenched lattice QCD calculations of $2600$ MeV this reduces to 
\be
a_\upmu^{G_{PS}(2600)}=0.19\times10^{-12}.
\ee
This is about an order of magnitude smaller than the pseudoscalar contribution
called $G/\eta''$ in the bottom-up holographic model of \cite{Leutgeb:2022lqw}, $a_\upmu^{\eta''}\approx 2\times 10^{-12}$.
In this more realistic model, the pseudoscalar glueball mixes not only with $\eta_0$ but also with excited $\eta(')$ mesons (which are absent in our simple extension of the WSS model to massive pseudoscalars).


\begin{acknowledgments}
We would like to thank Claude Amsler for useful discussions. We are also indebted to Jonas Mager for his assistance in the numerical
evaluation of the contributions to the anomalous magnetic moment of the muon.
F.~H.\ and J.~L.\ have been supported by the Austrian Science Fund FWF, project no. P 33655-N and the FWF doctoral program
Particles \& Interactions, project no. W1252-N27.
\end{acknowledgments}

\appendix

\section{Hadronic decays of the scalar and tensor glueballs}
\label{app:hadrGBdecays}

In the following we review the hadronic decays of scalar and tensor glueballs
in the WSS model as worked out in \cite{Brunner:2015oqa,Brunner:2015yha,Brunner:2015oga}, including additional
subdominant decay channels neglected therein, in particular $G\to a_1\pi$.
The latter has been emphasized in the phenomenological analysis of
\cite{Burakovsky:1998zg}, where it was providing the largest partial decay width
of a pure glueball (177 MeV for a glueball mass of 1600 MeV). While their results for decays of a scalar glueball into two vector mesons are remarkably compatible
with the WSS result for $G_D$ when the mass is raised to 1500-1700 MeV,
the WSS prediction for $G\to a_1\pi$ turns out
to be fairly small, $\lesssim 1$ MeV, in stark contrast to the model of
\cite{Burakovsky:1998zg}.\footnote{For $f_0(1500)$ the experimental value
from \cite{CRYSTALBARREL:2001ldq} is 12(5)\% of $\Gamma_{4\pi}$,
i.e., $\sim 7$ MeV; for $f_0(1710)$ no corresponding experimental results seem to be available.}

We also review the dependence on the so far unconstrained extra coupling
to be associated with the quark mass term that we have added to the chiral WSS model
(parametrized by $x$ in Table \ref{tab:GDEtotalwidth}).
As discussed in \cite{Brunner:2015oga}, this correlates the flavor asymmetries in the decay pattern in two pseudoscalars with the $\eta\eta'$ partial width.
Good agreement of the decay pattern of $G_D$ with $f_0(1710)$ (or $f_0(1770)$)
is obtained only for small or vanishing $\eta\eta'$ decay rates.
Here a new experimental result has been published in \cite{BESIII:2022iwi}:
$\mathcal{B}(f_0(1710)\to\eta\eta')/\mathcal{B}(f_0(1710)\to\pi\pi)<1.61\times 10^{-3}$, contradicting \cite{Sarantsev:2021ein,Klempt:2021wpg} where
this ratio is $\sim 1$ for $f_0(1710)$ and $\sim 0.1$ for $f_0(1770)$.

\subsection{Dilatonic scalar glueball}

The scalar glueball fluctuation which in \cite{Brunner:2015oqa}
is referred to as (predominantly) dilatonic scalar glueball,
reads 
\begin{align}
    \delta G_{\mu\nu} & =\frac{r^{2}}{\mathcal{N}_{D}\,L^{2}}T_{4}(r)\left(\eta_{\mu\nu}-\frac{\partial_{\mu}\partial_{\nu}}{\Box}\right)G_{D}(x^{\sigma}),\nonumber \\
    \delta G_{11,11} & =-3\frac{r^{2}}{\mathcal{N}_{D}\,L^{2}}T_{4}(r)G_{D}(x^{\sigma}),\label{eq:dilatonFluc}
\end{align}
with an undetermined normalization parameter $\mathcal{N}_{D}$. To
be a solution of the Einstein equations, the radial function $T_{4}(r)$
has to satisfy the differential equation
\begin{equation}\label{T4eq}
    \frac{\d}{\d r}\left(r^{7}-r\,r_{KK}^{6}\right)\frac{\d}{\d r}T_{4}(r)+L^{4}M_{D}^{2}r^{3}T_{4}(r)=0,
\end{equation}
with boundary conditions $T_{4}(r_{\kk})=1$ and $T_{4}^{\prime}(r_{\kk})=0$,
and therefore is normalizable for a discrete set of mass eigenvalues
$M_{D}$. In the following, we will only consider the lightest mode
with $M_{D}=1.567\MKK=1487\,\text{MeV}$.

The kinetic and mass term for $G_{D}$ reads
\begin{align}
    \left.\mathcal{L}_{4}\right|_{G_{D}^{2}} & =\mathcal{C}\int\text{d}r\frac{3r^{3}T_{4}(r)^{2}}{L^{3}\mathcal{N}_{D}^{2}}G_{D}\left(\Box-M_{D}^{2}\right)G_{D}
\end{align}
with the constant
\begin{align}
    \mathcal{C}= & \left(\frac{L}{2}\right)^{4}\Omega_{4}\frac{1}{2\kappa_{11}^{2}}\left(2\pi\right)^{2}R_{4}R_{11}.
\end{align}
The radial integration for the lightest mode yields the constant
\begin{align}
    \int\d r\frac{r^{3}T_{4}(r)^{2}}{L^{3}}=0.22547\left[T_{4}(r_{\kk})\right]^{2}\frac{r_{\kk}^{4}}{L^{3}}.\label{eq:rModeConstant-1-1}
\end{align}
To get a canonically normalized kinetic term 
\begin{align}
    \left.\mathcal{L}_{4}\right|_{G_{D}^{2}} & =\frac{1}{2}G_{D}\left(\Box-M_{D}^{2}\right)G_{D},
\end{align}
we have to set
\begin{equation}
    \mathcal{N}_{D}=0.0335879\lambda^{\frac{1}{2}}N_{C}M_{\kk}.
\end{equation}


Inducing the fluctuation \eqref{eq:dilatonFluc} in the D8 brane action
\eqref{eq:9dAction} we obtain the derivative coupling of two pseudoscalar mesons to $G_D$ as
\begin{equation}\label{Ld1}
    \mathcal{L}_{G_{D}\Pi\Pi}=d_1\tr\partial_\mu\Pi\partial_\nu\Pi\left(\eta^{\mu\nu}-\frac{\partial^\mu\partial^\nu}{\Box}\right)G_D
\end{equation}
where
\be
d_1=\frac{17.2261}{\sqrt{\lambda}M_\kk N_c}
\ee
(see \cite{Brunner:2015oqa} for further couplings).

Already in the chiral WSS model, a mass term arises for the singlet component
of $\Pi$ through the $U(1)_A$ anomaly \cite{Sakai:2004cn}.
The latter requires a redefinition of the Ramond-Ramond 2-form field strength $F_2$ which is associated with a $\theta$ term. The bulk action is thus given by
\begin{equation}
    S_{C_1}=-\frac{1}{4\pi(2\pi l_s)^6}\int\d^{10}x\sqrt{-g}|\Tilde{F}_2|^2,
\end{equation}
where
\begin{equation}
    \Tilde{F}_2=\frac{6\pi U_\kk}{U^4 M_\kk}\left(\theta +\frac{\sqrt{2 N_f}}{f_\pi}\eta_0\right)\d U\wedge dx^4,
\end{equation}
from which one obtains the Witten-Veneziano mass as \cite{Sakai:2004cn}
\begin{equation}
    m_0^2=\frac{N_f}{27\pi^2 N_c}\lambda^2 M_\kk.
\end{equation}

Inducing the metric fluctuations gives rise to an additional coupling between the scalar glueballs and $\eta_0$. For the dilatonic glueball it is given by \cite{Brunner:2015yha,Brunner:2015oga}
\begin{equation}
    \mathcal{L}_{\eta_0} \supset \frac{3}{2}m_0^2\eta_0^2d_0 G_D ,
\end{equation}
with ($H_D \equiv T_4/\mathcal{N}_D$)
\begin{equation}
    d_0=3U^3_\kk\int_{U_\kk}^\infty\d U H_D(U)U^{-4}\approx\frac{17.915}{\sqrt{\lambda}N_c M_\kk}.
\end{equation}

Massive quarks can be introduced by worldsheet instantons \cite{Aharony:2008an,Hashimoto:2008sr,Bergman:2007pm} or tachyon condensation \cite{Dhar:2007bz,Dhar:2008um,McNees:2008km}, which give
\begin{equation}
    \mathcal{L}_m^{\mathcal{M}}\propto \int\d^4x\mathrm{Tr}\left(\mathcal{M}U(x)+h.c.\right),
\end{equation}
where
\begin{equation}
    U(x)=\mathrm{P}\exp i\int\d z A_z(z,x)=e^{i\Pi^a\lambda^a/f_\pi}.
\end{equation}

Expanding the mass term with $\mathcal{M}=\mathrm{diag}(\hat{m},\hat{m},m_s)$ leads to
\begin{align}
     \mathcal{L}_m^{\mathcal{M}}=-\frac{1}{2}m_\pi^2\pi_0^2-m_\pi^2\pi^+\pi^--m_K^2(K_0\Bar{K}_0+K_+K_-)\nonumber\\
     -\frac{1}{2}m_1^2\eta_0^2-\frac{1}{2}m_8^2\eta_8+\frac{2\sqrt{2}}{3}(m_K^2-m_\pi^2)\eta_0\eta_8,
\end{align}
with
\begin{align}
    m_\pi^2=2\hat{m}\mu,\quad m_K^2=(\hat{m}+m_s)\mu,\nonumber\\
    m_1^2=\frac{2}{3}m_K^2+\frac{1}{3}m_\pi^2,\quad m_8^2=\frac{4}{3}m_K^2-\frac{1}{3}m_\pi^2,
\end{align}
and $\mu$ being the overall scale. We also note a sign error in the $\eta_0\eta_8$ mixing term in \cite{Brunner:2015yha}. With
\begin{equation}
    \eta=\eta_8 \cos\theta_P-\eta_0\sin\theta_P,\quad \eta^\prime=\eta_8\sin\theta_P+\eta_0\cos\theta_P
\end{equation}
the mass term is diagonalized by
\begin{equation}
    \theta_P=\frac{1}{2}\arctan\frac{2\sqrt{2}}{1-\frac{3}{2}m_0^2/(m_K^2-m_\pi^2)}
\end{equation}
leading to
\begin{equation}
    m_{\eta,\eta^\prime}^2=\frac{1}{2}m_0^2+m_K^2\mp\sqrt{\frac{m_0^4}{4}-\frac{1}{3}m_0^2(m_K^2-m_\pi^2)+(m_K^2-m_\pi^2)^2}
\end{equation}
for the $\eta$ and $\eta^\prime$ meson, respectively. 

As in \cite{Brunner:2015yha,Brunner:2015oga}, we assume a scalar glueball coupling to the quark mass terms of the form (correcting a typo in \cite{Brunner:2015oga})
\begin{equation}\label{GDmassterm}
    \mathcal{L}_{G_Dq\Bar{q}}=-3d_m G_D \mathcal{L}_m^{\mathcal{M}}
\end{equation}
with $d_m$ being of the same order as $d_0$, i.e.
\begin{equation}
    d_m=x d_0,\quad x=\mathcal{O}(1).
\end{equation}
This leads to a $G_D\eta\eta^\prime$ interaction given by
\begin{equation}
    \mathcal{L}_{G_D\eta\eta^\prime}=-\frac{3}{2}(1-x)d_0\sin (2\theta_P) m_0^2G_D\eta\eta^\prime.
\end{equation}

With these modifications we obtain the coupling of the dilaton glueball to $\eta\eta$ as
\begin{equation}
    \mathcal{L}_{G_D\eta\eta}=\frac{3}{2}d_0 m_0^2 (1-x)\sin\theta^2_P G_D\eta\eta+\frac{3}{2}d_0 x m_\eta^2 G_D\eta\eta+\frac{d_1}{2}\partial_\mu\eta\partial_\nu\eta\left(\eta^{\mu\nu}-\frac{\partial^\mu\partial^\mu}{\Box}\right)G_D .
\end{equation}
For the coupling to the $\eta^\prime$ meson we get $\cos\theta_P^2$ instead of $\sin\theta_P^2$. 

The partial decay width for $G_D$ decaying into two identical pseudoscalar mesons becomes
\begin{equation}
    \Gamma_{G_D\to PP}=\frac{n_P d_1^2 M_D^3}{512\pi}\left(1-4\frac{m_P^2}{M_D^2}\right)^{1/2}\left(1+\alpha\frac{m_P^2}{M_D^2}\right)^2,
\end{equation}
where $P$ refers to pions ($n_P=3$), kaons ($n_P=4$) or $\eta^{(')}$ ($n_P=1)$ mesons, and
\begin{equation}
    \alpha=4\left(3\frac{d_0}{d_1}x-1\right)
\end{equation}
for pions and kaons, and
\begin{equation}
    \alpha=4\left[3\frac{d_0}{d_1}\left(x+\frac{m_0^2}{m_P^2}\sin^2\theta_P(1-x)\right)-1\right].
\end{equation}
for  $\eta\eta$, and with the replacement $\sin\theta_P\to\cos\theta_P$ for $\eta^\prime\eta^\prime$.


There is also a trilinear coupling of a dilatonic scalar glueball with
one axial vector and one pseudoscalar meson, which has been
neglected in \cite{Brunner:2015oqa}, given by
\begin{align}
    \mathcal{L}_{G_{D} \Pi a^{m}}= & -2d_{6}^{m } M_\kk\tr \left(\partial_\mu \Pi a_\nu^{m}  \right) \left(\eta^{\mu\nu}-\frac{\partial^{\mu}\partial^{\nu}}{\Box}\right)G_{D},
\end{align}
with 
\begin{align}
    d_{6}^{m} & \equiv  \sqrt{\frac{\kappa}{\pi}} \int \d z\, \psi_{2m}^\prime H_{D}\nonumber\\
    & = \left\{11.768, 7.809,2.350,\ldots
    \right\} \frac{1}{M_{\kk}N_c\sqrt{\lambda}}.
\end{align}

Restricting ourselves to two-body decays, for which the relevant vertices for
vector mesons are given in Sect.\ \ref{sec:GDdecays}, 
the resulting partial decay widths
are collected in Table \ref{tab:GDhadronicdecays}.

\begin{table}
    \centering{}\bigskip{}
    \begin{tabular}{lcccc}
    \toprule
    & $\Gamma_{G_{D}}^{WSS}${[}MeV{]}  & $\Gamma_{G_{D}(1506)}${[}MeV{]} & $\Gamma_{G_{D}(1712)}${[}MeV{]} & $\Gamma_{G_{D}(1865)}${[}MeV{]} \tabularnewline\colrule
    $G_{D}\rightarrow \pi\pi $ & 12.4\dots 16.5$|$15.2\dots 20.1 & 12.6\dots 16.7$|$15.4\dots 20.4  & 14.6\dots 19.3$|$17.0\dots 22.5  &  16.1\dots 21.3$|$18.3\dots 24.2 \tabularnewline
    $G_{D}\rightarrow KK $ & 4.16\dots 5.51$|$50.5\dots 67.0 &  4.43\dots 5.87$|$50.4\dots 66.8 & 7.49\dots 9.93$|$49.4\dots 65.4 & 9.87\dots 13.1$|$48.8\dots 64.7 \tabularnewline
    $G_{D}\rightarrow \eta\eta$ & 1.85\dots 3.71$|$14.1\dots 18.7 & 1.93\dots 3.82$|$14.1\dots 18.7 & 2.77\dots 4.96$|$13.9\dots 18.4 & 3.38\dots 5.75$|$13.7\dots 18.1\tabularnewline
    $G_{D}\rightarrow \eta\eta^\prime$ & - & 0.29\dots 0.30$|$0\phantom{.00\dots0.00}   & 4.35\dots 4.54$|$0\phantom{.00\dots0.00} & 4.19\dots 4.38$|$0\phantom{.00\dots0.00} \tabularnewline
    $G_{D}\rightarrow a_1\pi$ & 0.14\dots 0.18 & 0.17\dots 0.23 & 0.66\dots 0.87 & 1.08\dots 1.43 \tabularnewline
    $G_{D}\rightarrow \rho\rho$ &-  & - & 53.5\dots 71.0 & 90.1\dots 119 \tabularnewline
    $G_{D}\rightarrow \omega\omega$ & - & - & 16.6\dots 22.0 &  28.7\dots 38.1 \tabularnewline
    $G_{D}\rightarrow K^*K^*$ & - & - & - & 42.6\dots 56.4 \tabularnewline\hline
    Sum & 18.6\dots 25.9$|$79.9\dots 106 & 19.4\dots 26.9$|$80.0\dots 106  & 100\dots 133$|$151\dots 200  & 196\dots 260$|$243\dots 322 \tabularnewline\botrule 
    \end{tabular}
    \caption{Hadronic two-body decays of the dilatonic scalar glueball $G_D$ with WSS model mass and extrapolated to the masses of $f_0(1500)$, $f_0(1710)$, and $M=1865$ MeV, for $\lambda=16.63\dots12.55$. In decays into two pseudoscalar mesons, the two sets of values correspond to $x=0$ and $x=1$ in the coupling to the quark mass term \eqref{GDmassterm}. 
    \label{tab:GDhadronicdecays}}
\end{table}

\subsection{Exotic scalar glueball}

The lighter exotic scalar glueball fluctuation
with mass $M_E=0.901\MKK=855$ MeV, which we have discarded from the spectrum
when identifying the dilatonic scalar glueball with the ground-state glueball of QCD, 
is given by \eqref{GEmetricfluctuations}
with eigenvalue equation \eqref{GEmodeeq}.
This mode involves the metric component $h_{\tau\tau}$, which has
no analogues in other holographic QCD models, and has therefore been
termed ``exotic'' in \cite{Constable:1999gb}. Its canonical normalization is
obtained from
\begin{align}
    \left.\mathcal{L}_{4}\right|_{G_{E}^{2}} & =\mathcal{C}\int\d r\frac{r^{3}S_{4}(r){}^{2}}{2L^{3}\mathcal{N}_{E}^{2}} \frac{5}{8} G_{E}\left(\Box-M_{E}^{2}\right)G_{E}\nonumber \\
     & =\frac{1}{2}G_{E}\left(\Box-M_{E}^{2}\right)G_{E},
\end{align}
with
\begin{align}
    \int\d r\frac{r^{3}S_{4}(r)^{2}}{L^{3}}=
    0.09183\left[S_{4}(r_{\kk})\right]^{2}\frac{r_{\kk}^{4}}{L^{3}}\label{eq:rModeConstant-1-4}
\end{align}
and
\be
\mathcal{N}_{E}=0.008751\lambda^{\frac{1}{2}}N_{C}M_{\kk}.
\ee

Derivative couplings of pseudoscalars to $G_E$ are given by \begin{equation}
    \mathcal{L}_{G_{E}}\supset -\tr\left\{c_1\left[\partial_\mu\Pi\partial_\nu\Pi\frac{\partial^\mu\partial^\nu}{M_E^2}G_E+\frac{1}{2}\left(\partial_\mu\Pi\right)^2\left(1-\frac{\Box}{M_E^2}\right)G_E\right]+\Breve{c}_1\partial_\mu\Pi\partial^\mu\Pi G_E\right\}
\end{equation}
with $c_1$ and $\breve{c}_1$ as in \cite{Brunner:2015oqa}.

In the Witten-Veneziano mass term for $\eta_0^2$, inducing the metric fluctuations leads to additional couplings between the scalar glueballs and $\eta_0$. For the exotic scalar glueball it is given by
\begin{equation}
    \mathcal{L}_{\eta_0} \supset -\frac{5}{2}m_0^2\eta_0^2\Breve{c}_0 G_E ,
\end{equation}
with ($H_E \equiv S_4/\mathcal{N}_E$)
\begin{equation}
  \Breve{c}_0=\frac{3}{4}U^3_\kk\int_{U_\kk}^\infty\d U H_E(U)U^{-4}\approx\frac{15.829}{\sqrt{\lambda}N_c M_\kk}
\end{equation}
as previously studied in \cite{Brunner:2015oga}.

Assuming the coupling of the exotic scalar glueball to quark masses to be of the form
\begin{equation}\label{GEmassterm}
    \mathcal{L}_{G_Eq\Bar{q}}=5\Breve{c}_m G_E \mathcal{L}_m^{\mathcal{M}}
\end{equation}
with $\Breve{c}_m$ being of the same order as $\Breve{c}_0$, i.e.
\begin{equation}
    \Breve{c}_m=x \Breve{c}_0,\quad x=\mathcal{O}(1),
\end{equation}
we get
\begin{equation}
    \mathcal{L}_{G_E\eta\eta^\prime}=\frac{5}{2}(1-x)\Breve{c}_0\sin (2\theta_P) m_0^2G_E\eta\eta^\prime.
\end{equation}

All together we obtain the coupling of the exotic scalar glueball to $\eta\eta$ as
\begin{align}
    \mathcal{L}_{G_E\eta\eta}=\frac{5}{2}\Breve{c}_0 m_0^2 (x-1)\sin\theta^2_P G_E\eta\eta-\frac{5}{2}\Breve{c}_0 x m_\eta^2 G_E\eta\eta\nonumber\\
    -\frac{c_1}{2}\partial_\mu\eta\partial_\nu\eta\left(\frac{1}{2}\eta^{\mu\nu}\left(1-\frac{\Box}{M_E^2}\right)+\frac{\partial^\mu\partial^\nu}{M_E^2}\right)G_E - \frac{\Breve{c}_1}{2}\partial_\mu\eta\partial^\mu\eta G_E.
\end{align}

For pions and kaons we have
\begin{equation}
    \left|\mathcal{M}_{G_E\to PP}\right|=\frac{1}{4}\left|20\Breve{c}_0 m_P^2 x + 2\Breve{c}_1(M_E^2-2m_P^2)+c_1 M_E^2\right|
\end{equation}
and for $\eta$
\begin{equation}
    \left|\mathcal{M}_{G_E\to\eta\eta}\right|=\frac{1}{4}\left|-20\Breve{c}_0 m_0^2(x-1)\sin\theta_P^2+20\Breve{c}_0 m_P^2 x+2\Breve{c}_1(M_E^2-2m_P^2)+c_1 M_E^2\right|,
\end{equation}
from which the $\eta'$ amplitude is obtained by the replacement $\sin\theta_P\to\cos\theta_P$.

In both cases the decay width is given by
\begin{equation}
     \Gamma_{G_E\to PP}=\frac{n_P}{2}\frac{1}{8\pi}\left|\mathcal{M}_{G_E\to PP}\right|^2\frac{\left|\mathbf{p}_{P}\right|}{M_E^2}.
\end{equation}

The coupling of the exotic scalar glueball
to one axial vector meson and one pseudoscalar meson is given by
\begin{align}
    \mathcal{L}_{G_{E} \Pi a^{m}}= & 2c_{6}^{m } M_\kk\tr \left(\partial_\mu \Pi a_\nu^{m}  \right)\frac{\partial^{\mu}\partial^{\nu}}{M_E^2}G_{E},
\end{align}
with 
\begin{align}
    c_{6}^{m} & =  \sqrt{\frac{\kappa}{\pi}} \int \d z\, \psi_{2m}^\prime \left[\frac{1}{4}+\frac{3}{5K-2}\right] H_{E}\nonumber\\
    & = \left\{57.659,72.057,65.190,\ldots\right\} \frac{1}{M_{\kk}N_c\sqrt{\lambda}}.
\end{align}

Restricting ourselves to two-body decays, for which the relevant vertices for
vector mesons are given separately in Appendix \ref{app:GE}, 
the resulting partial decay widths
are collected in Table \ref{tab:GEhadronicdecays}.

\begin{table}
    \centering{}\bigskip{}
    \begin{tabular}{lcccc}
    \toprule
    & $\Gamma_{G_{E}}^{WSS}${[}MeV{]}  & $\Gamma_{G_{E}(1506)}${[}MeV{]} & $\Gamma_{G_{E}(1712)}${[}MeV{]} & $\Gamma_{G_{E}(1865)}${[}MeV{]} \tabularnewline\colrule
    $G_{E}\rightarrow \pi\pi $ &  72.2\dots 95.7$|$84.9\dots 113 & 135\dots 179$|$142\dots 189  & 154\dots 205$|$161\dots 213  &    169\dots 224$|$175\dots 231  \tabularnewline
    $G_{E}\rightarrow KK $ & - &  120\dots 158$|$229\dots 304 & 152\dots 202$|$255\dots 338 & 176\dots 233$|$273\dots 362 \tabularnewline
    $G_{E}\rightarrow \eta\eta$ & - & 31.3\dots 45.4$|$57.7\dots 76.4   & 40.0\dots 56.9$|$65.1\dots 86.3 & 45.9\dots 64.6$|$69.8\dots 92.5\tabularnewline
    $G_{E}\rightarrow \eta\eta^\prime$ & - & 0.21\dots 0.22$|$0\phantom{.00\dots0.00}   & 3.12\dots 3.26$|$0\phantom{.00\dots0.00} & 3.01\dots 3.14$|$0\phantom{.00\dots0.00}\tabularnewline
    $G_{E}\rightarrow a_1\pi$ & - & 0.06\dots 0.08 & 0.55\dots 0.73 & 1.36\dots 1.80 \tabularnewline
    $G_{E}\rightarrow \rho\rho$ &-  & - & 0.77\dots 1.02 & 2.91\dots 3.86 \tabularnewline
    $G_{E}\rightarrow \omega\omega$ & - & - & 0.19\dots 0.26 & 0.84\dots 1.12\tabularnewline
    $G_{E}\rightarrow K^*K^*$ & - & - & - & 0.15\dots 0.20\tabularnewline\hline
    Sum & 72.2\dots 95.7$|$84.9\dots 113 & 286\dots 383$|$430\dots 570  & 351\dots 469$|$483\dots 640  & 398\dots 530$|$521\dots 691 \tabularnewline\botrule 
    \end{tabular}
    \caption{Hadronic two-body decays of the exotic scalar glueball $G_E$ with WSS model mass 855 MeV and extrapolated to the masses of $f_0(1500)$, $f_0(1710)$, and the scalar glueball at 1865 MeV proposed in \cite{Sarantsev:2021ein}, for $\lambda=16.63\dots12.55$. In decays into two pseudoscalar mesons, the two sets of values correspond to $x=0$ and $x=1$ in the coupling to the quark mass term \eqref{GEmassterm}. 
    \label{tab:GEhadronicdecays}}
\end{table}

\subsection{Tensor glueball}\label{app:tensor}

The tensor glueball fluctuations read 
\begin{equation}
    h_{\mu\nu}=q{}_{\mu\nu}\frac{r^{2}}{L^{2}\,\mathcal{N}_{T}}T_4(r)G_{T}(x^{\sigma}),
\end{equation}
where $q_{\mu\nu}$ is a symmetric, transverse, and traceless polarization
tensor, which we normalize such that $q_{\mu\nu}q^{\mu\nu}=1$, differing
from \BPR. 

$T_{4}(r)$ satisfies the same eigenvalue equation as in the case of the dilatonic
scalar glueball, \eqref{T4eq}, but it acquires a different normalization.
The Lagrangian reads 
\begin{align}
    \left.\mathcal{L}_{4}\right|_{G_{T}^{2}} &= \mathcal{C}\int\d r\frac{r^{3}T_{4}(r)^{2}}{4L^{3}\mathcal{N}_{T}^{2}}G_{T}\left(\Box-M^{2}\right)G_{T}\nonumber \\
    &= \frac{1}{2}G_{T}\left(\Box-M^{2}\right)G_{T},
\end{align}
with 
\begin{align}
    \int\d r\frac{r^{3}T_4(r)^{2}}{2L^{3}}=0.112735 [T_4(r_{\kk})]^{2}\frac{r_{\kk}^{4}}{L^{3}}
\end{align}
and 
\begin{equation}
    \mathcal{N}_{T}=0.00969598\lambda^{\frac{1}{2}}N_{C}M_{\kk}=\frac{1}{2\sqrt{3}}\mathcal{N}_{D}.
\end{equation}

This leads to
\begin{equation}
    \mathcal{L}_{G_T\Pi\Pi}=t_1\tr\left(\partial_\mu\Pi\partial_\nu\Pi\right)G^{\mu\nu}_T
\end{equation}
with ($T\equiv T_4/\mathcal{N}_T$)
\begin{equation}
    t_1=\frac{1}{\pi}\int\d zK^{-1} T=\frac{59.6729}{\sqrt{\lambda}M_\kk N_c}=2\sqrt{3}d_1.
\end{equation}
Here no additional couplings arise from the mass terms of the pseudoscalars,
because the tensor glueball fluctuations are traceless.

There is also a coupling of the tensor glueball to one axial vector and
one pseudoscalar meson,
\be
    \mathcal{L}_{G_{T} \Pi a^{m}}= -2t_{6}M_{\kk} \tr\left( \partial_{\mu}\Pi a_{\nu}^m\right)G_{T}^{\mu\nu}
\ee
with
\begin{equation}
    t_{6}=\sqrt{\frac{\kappa}{\pi}}\int\d z \psi_{2m}^\prime T=\left\{40.764,27.050,8.140,\ldots\right\} \frac{1}{M_{\kk}N_c\sqrt{\lambda}}.
\end{equation}

Restricting ourselves to two-body decays, for which the relevant vertices for
vector mesons are given in Sect.\ \ref{sec:GTdecays}, 
the resulting partial decay widths
are collected in Table \ref{tab:GThadronicdecays}.

\begin{table}
    \centering{}\bigskip{}
    \begin{tabular}{lccc}
    \toprule
    & $\Gamma_{G_{T}}^{WSS}${[}MeV{]}  & $\Gamma_{G_{T}(2000)}${[}MeV{]} & $\Gamma_{G_{T}(2400)}${[}MeV{]} \tabularnewline\colrule
    $G_{T}\rightarrow \pi\pi $ & 19.9\dots 26.3 & 27.7\dots 36.8   &  33.8\dots 44.7  \tabularnewline
    $G_{T}\rightarrow KK $ & 6.66\dots 8.83 & 19.2\dots 25.4  & 29.2\dots 38.6\tabularnewline
    $G_{T}\rightarrow \eta\eta$ & 1.02\dots 1.35  &  3.97\dots 5.26 & 6.48\dots 8.58 \tabularnewline
    $G_{T}\rightarrow a_1\pi$ & 0.53\dots 0.71 & 5.12\dots 6.78 & 8.00\dots 10.6 \tabularnewline
    $G_{T}\rightarrow\rho\rho$   & - & 270\dots 358 & 382\dots 507 \tabularnewline
    $G_{T}\rightarrow\omega\omega$  & - & 88.2\dots 117 & 127\dots 169 \tabularnewline 
    $G_{T}\rightarrow K^{*}K^{*}$  & - & 240\dots 318 & 417\dots 552 \tabularnewline
    $G_{T}\rightarrow f_1\eta$ & - & 0.98\dots 
    1.71 & 3.97\dots 
    6.89\tabularnewline
    $G_{T}\rightarrow \eta^\prime\eta^\prime$ & -  & - 
    & 0.92\dots 1.22\tabularnewline
    $G_{T}\rightarrow\phi\phi$  & - & - & 76.7\dots 102 \tabularnewline
    \hline
    Total & 28.1\dots 37.2 & 655\dots 869 & 1084\dots 1437 \tabularnewline
    \botrule 
    \end{tabular}
    \caption{Hadronic two-body decays of the tensor glueball $G_{T}$ with WSS model 1487 MeV mass and extrapolated to masses of 2000 and 2400 MeV, for $\lambda=16.63\dots12.55$. In decays involving $f_1$ we additionally vary $\theta_{f}=20.4^\circ\dots26.4^\circ$. Partial decay widths much smaller than 1 MeV are left out.
    \label{tab:GThadronicdecays}}
\end{table}

{
Recently, Ref.~\cite{Vereijken:2023jor} calculated branching ratios of tensor glueball decays in a chiral hadronic model, the so-called extended linear sigma model, where the ratios of all the decay modes of Table \ref{tab:GThadronicdecays} can be obtained, although not their absolute magnitudes. In that model a similar dominance of decays into two vector mesons (when kinematically allowed) has been obtained, which is numerically even more pronounced.\footnote{For example, while in the WSS model the branching ratio $\rho\rho:\pi\pi$ is around 10-11 for a tensor glueball mass between 2000 and 2400 MeV, in Ref.~\cite{Vereijken:2023jor} it varies between 60 and 50. Also the branching ratio $\rho\rho:a_1\pi$ is 6 to 5 times larger there for this mass range.} The authors of Ref.~\cite{Vereijken:2023jor} also gave a rough estimate of $\Gamma(G_T\to\pi\pi)\sim 15$ MeV, which turns out to be comparable with the WSS result.
}

\section{Radiative Decays of the Exotic Scalar Glueball}
\label{app:GE}

\begin{widetext}
The exotic glueball interactions contain the vertices
    \begin{align}\label{LGEvv}
    \mathcal{L}_{G_{E} v^{m} v^{n}}= & -\tr\left\{ c_{2}^{mn}M_{\kk}^{2}\left[ v_{\mu}^{m} v_{\nu}^{n}\frac{\partial^{\mu}\partial^{\nu}}{M_{E}^{2}}G_{E}+\frac{1}{2} v_{\mu}^{m} v^{n\mu}\left(1-\frac{\Box}{M_{E}^{2}}\right)G_{E}\right]\right.\nonumber \\
    & +c_{3}^{mn}\left[\eta^{\rho\sigma}F_{\mu\rho}^{m}F_{\nu\sigma}^{n}\frac{\partial^{\mu}\partial^{\nu}}{M_{E}^{2}}G_{E}-\frac{1}{4}F_{\mu\nu}^{m}F^{n\mu\nu}\left(1+\frac{\Box}{M_{E}^{2}}\right)G_{E}\right]+3c_{4}^{mn}\frac{M_{\kk}^{2}}{M_{E}^{2}} v_{\mu}^{n}F^{m\mu\nu}\partial_{\nu}G_{E}\nonumber \\
    & \left.+\breve{c}_{2}^{mn}M_{\kk}^{2} v_{\mu}^{m} v^{n\mu}G_{E}+\frac{1}{2}\breve{c}_{3}^{mn}F_{\mu\nu}^{m}F^{n\mu\nu}G_{E}\vphantom{\left(\frac{\Box}{M_{E}^{2}}\right)}\right\} ,
\end{align}    
\end{widetext}

with coupling constants 
\begin{align}\label{c234}
    &c_{2}^{mn}= \kappa\int\d zK\psi_{2m-1}^{\prime}\psi_{2n-1}^{\prime}\overline{H}_{E}=\frac{\{7.116,\ldots\}}{\MKK N_c \sqrt{\lambda}},\nonumber\\ 
    &c_{3}^{mn}=\kappa\int\d zK^{-1/3}\psi_{2m-1}\psi_{2n-1}\overline{H}_{E}=\frac{\{69.769,\ldots\}}{\MKK N_c \sqrt{\lambda}}\nonumber \\
    &c_{4}^{mn}= \kappa \int\d z\,\frac{20zK}{\left(5K-2\right)^{2}}\psi_{2m-1}\psi_{2n-1}^{\prime}H_{E}=\frac{\{-10.5798,\ldots\}}{M_{\kk}N_c \sqrt{\lambda}},\nonumber\\
    &\breve{c}_{2}^{mn}=\frac{\kappa}{4}\int\d zK\psi_{2m-1}^{\prime}\psi_{2n-1}^{\prime}H_{E}=\frac{\{2.966,\ldots\}}{\MKK N_c \sqrt{\lambda}}\nonumber \\
    &\breve{c}_{3}^{mn}=\frac{\kappa}{4}\int\d zK^{-1/3}\psi_{2m-1}\psi_{2n-1}H_{E}=\frac{\{18.122,\ldots\}}{\MKK N_c \sqrt{\lambda}},
\end{align}
where $\overline{H}_{E}= \left[\frac{1}{4}+\frac{3}{5K-2}\right]H_{E}$.

Calculating the amplitude for different polarizations we get 
\begin{align}
    \left|\mathcal{M}_{T}^{\left(G_{E}\rightarrow v^{1}v^{1}\right)}\right| & =\frac{1}{2}\left[c_{3}\left(M_E^{2}-4m_{v}^{2}\right)-6c_{4}\MKK^2-4\breve{c}_{2}M_{\kk}^{2}-2\breve{c}_{3}\left(M_E^{2}-2m_{v}^{2}\right)\right] \nonumber \\
    \left|\mathcal{M}_{L}^{\left(G_{E}\rightarrow v^{1}v^{1}\right)}\right| & =\frac{c_{2}M_{\kk}^{2}\left(M_E^{2}-4m_{v}^{2}\right)+2\breve{c}_{2}M_{\kk}^{2}\left(M_E^{2}-2m_{v}^{2}\right)+6c_{4}\MKK^2 m_{v}^{2}+4\breve{c}_{3}m_{v}^{4}}{2m_{v}^{2}} .
\end{align}

\subsubsection{Exotic scalar glueball \texorpdfstring{$1\gamma$-decays}{1 photon}}

For the decay in one vector meson and one photon, we use 
\begin{widetext}
\begin{align}
    \mathcal{L}_{G_{E}\V v^{m}}=-\tr & \left\{ c_{3}^{m\mathcal{V}}\left[2\eta^{\rho\sigma}F_{\mu\rho}^{m}F_{\nu\sigma}^{\V}\frac{\partial^{\mu}\partial^{\nu}}{M_{E}^{2}}G_{E}-\frac{1}{2}F_{\mu\nu}^{m}F^{\mathcal{V}\mu\nu}\left(1+\frac{\Box}{M_{E}^{2}}\right)G_{E}\right]\right.\nonumber \\
     & +3c_{4}^{\mathcal{V}n}\frac{M_{\kk}^{2}}{M_{E}^{2}} v_{\mu}^{n}F^{\mathcal{V}\mu\nu}\partial_{\nu}G_{E}\left.+\breve{c}_{3}^{m\mathcal{V}}F_{\mu\nu}^{m}F^{\V\mu\nu}G_{E}\vphantom{\left(\frac{\Box}{M_{E}^{2}}\right)}\right\} ,
\end{align}
with 
\begin{align}
    c_{3}^{m\mathcal{V}}= &\kappa \int\d zK^{-1/3}\psi_{2m-1}\overline{H}_{E}
     = \frac{\{ 1.551,\ldots\}}{M_{\kk}N_{c}^{\frac{1}{2}}}\nonumber\\ 
    c_{4}^{\mathcal{V}m}=&\kappa \int\d z\,\frac{20ZK}{\left(5K-2\right)^{2}}\psi_{2m-1}^{\prime}H_{E}\nonumber
    = \frac{\{ -0.262,\ldots\}}{M_{\kk}N_{c}^{\frac{1}{2}}}\nonumber\\
    \breve{c}_{3}^{m\mathcal{V}}= &\frac{ \kappa}{4}\int\d zK^{-1/3}\psi_{2m-1}H_{E},\nonumber
    =  \frac{\{ 0.425,\ldots\}}{M_{\kk}N_{c}^{\frac{1}{2}}}
\end{align}
to obtain 
\begin{align}
    \left|\mathcal{M}_{T}^{\left(G_{E}\rightarrow v^{m}\V\right)}\right| & =\frac{\left(M_E^{2}-m_{v}^{2}\right)}{2M_E^{2}}\left|3c_{4}^{\V n}\MKK^2+2\breve{c}_{3}^{m\V}M_E^{2}+c_{3}^{m\V}\left(m_{v}^{2}-M_E^{2}\right)\right|  \tr\left( e Q T_{v^m }\right).
\end{align}
\end{widetext}

\subsubsection{Exotic scalar glueball \texorpdfstring{$2\gamma$-decays}{2 photon}}

The two-photon decay rate is obtained from
\begin{align}
    \mathcal{L}_{G_{E}\V\V}= & -\tr\left\{ c_{3}^{\V\V}\left[F_{\mu\rho}^{\mathcal{\V}}F_{\nu}^{\V\rho}\frac{\partial^{\mu}\partial^{\nu}}{M_{E}^{2}}G_{E}-\frac{1}{4}F_{\mu\nu}^{\V}F^{\V\mu\nu}\left(1+\frac{\Box}{M_{E}^{2}}\right)G_{E}\right]\right.\nonumber \\
    & \left.+\frac{1}{2}\breve{c}_{3}^{\V\V}F_{\mu\nu}^{\V}F^{\V\mu\nu}G_{E}\vphantom{\left(\frac{\Box}{M_{E}^{2}}\right)}\right\} 
\end{align}
with 
\begin{align}
    c_{3}^{\V\V}&= \kappa\int\d zK^{-1/3}\overline{H}_{E}=\frac{237.587\kappa}{M_{\kk}N_{c}\lambda^{1/2}}=0.0355\frac{\lambda^{\frac{1}{2}}}{M_{\kk}},\\
    \breve{c}_{3}^{\V\V}&= \frac{\kappa}{4}\int\d zK^{-1/3}H_{E}=\frac{71.18\kappa}{M_{\kk}N_{c}\lambda^{1/2}}=0.0106\frac{\lambda^{\frac{1}{2}}}{M_{\kk}},
\end{align}
yielding
\be
    \left|\mathcal{M}_{T}^{\left(G_{E}\rightarrow \V\V\right)}\right|  =\frac{M_E^{2}}{2}\left(c_{3}^{\mathcal{V}\mathcal{V}}-2\breve{c}_{3}^{\mathcal{V}\mathcal{V}}\right) \tr \left( e^2 Q^2\right).
\ee

In Table \ref{radGEDecay} the results for the partial widths for the
radiative and two-vector decays of the exotic scalar glueball are
given when the above amplitudes are substituted in the respective formulae
for the dilaton scalar glueball, \eqref{GammaGDvv}, \eqref{GammaGDvV},
and \eqref{GammaGDVV}.
Again, these are evaluate for the WSS model mass, which is only 855 MeV for the exotic
scalar glueball, as well as for three higher masses, corresponding to
the glueball candidates $f_0(1500)$, $f_0(1710)$, and the
one proposed in \cite{Sarantsev:2021ein}.
While the total decay width of $G_E$ is much larger than that of $G_D$
at equal mass, see Table \ref{tab:GDEtotalwidth}, the radiative and
two-vector widths of $G_E$ are much smaller than those of $G_D$, see \ref{tab:radGDDecayideal}.

\begin{table}
    \centering{}\bigskip{}
    \begin{tabular}{lcccc}
    \toprule
    & $\Gamma_{G_{E}^\mathrm{WSS}}${[}keV{]} & $\Gamma_{G_{E}(1506)}${[}keV{]} & $\Gamma_{G_{E}(1712)}${[}keV{]} & $\Gamma_{G_{E}(1865)}${[}keV{]}\tabularnewline\colrule   
    $G_{E}\rightarrow \rho\rho$ & - & - & 771\dots 1022 & 2910\dots 3857\tabularnewline
    $G_{E}\rightarrow \omega\omega$ & - & - & 194\dots 257 & 843\dots 1117\tabularnewline
    $G_{E}\rightarrow K^*K^*$ & - & - & - & 149\dots 197\tabularnewline\colrule
    $G_{E}\rightarrow \rho\gamma$    & 0.047 & 13.4 &  20.7 & 26.4\tabularnewline
    $G_{E}\rightarrow \omega\gamma$    & 0.003 & 1.4  & 2.23 & 2.86 \tabularnewline
    $G_{E}\rightarrow \phi\gamma$    & - & 0.30 & 0.98 & 1.72 \tabularnewline\colrule  
    $G_{E}\rightarrow \gamma\gamma$  & 0.043\dots 0.033 & 0.076\dots0.058 & 0.087\dots0.066 & 0.095\dots 0.071\tabularnewline\botrule 
    \end{tabular}
    \caption{Radiative and two-vector decays of the exotic scalar glueball $G_E$ with WSS model mass 855 MeV and extrapolated to the masses of $f_0(1500)$, $f_0(1710)$ and the scalar glueball at 1865 MeV proposed in \cite{Sarantsev:2021ein}. 
    \label{radGEDecay} }
\end{table}

\newpage

    \bibliographystyle{bib/JHEP}
    \bibliography{bib/references}

\end{document}